\documentclass[manuscript]{emulateapj}

\usepackage{graphicx}
\newcommand{\dxdy}[2]{{\frac{\partial{#1}}{\partial{#2}}}}

\newcommand{\DxDy}

\shorttitle{Internal gravity waves in massive stars}
\shortauthors{Rogers et al.}

\begin{document}







\title{Internal Gravity Waves in Massive Stars: Angular Momentum
  Transport}
\author{T.M. Rogers} \affil{Department of Planetary Sciences,
  University of Arizona, Tucson, AZ, 85719}
\email{tami@lpl.arizona.edu} \author{D.N.C. Lin} \affil{Astronomy and
  Astrophysics Department, University of California, Santa Cruz, CA
  95064} \affil{Kavli Institute for Astronomy and Astrophysics and
  School of Physics, Peking University, China} \affil{Institute for
  Advanced Studies, Tsinghua University, Beijing, China} \email{lin@ucolick.org}
\author{J.N. McElwaine} \affil{Swiss Federal Institute for Snow and
  Avalanche Research, 11 Fluelastrasse, Davos Dorf, Switzerland}
\affil{Planetary Science Institute, Tucson AZ 85721}
\email{james.mcelwaine@slf.ch} \author{H.H.B. Lau}
\affil{Argelander-Institut for Astronomie Universit Bonn Auf dem
  Huegel 71 53121 Bonn Germany} \affil{Monash Centre for Astrophysics,
  School of Mathematical Sciences, Monash University, Australia}
\email{hblau@astro.uni-bonn.de}


\begin{abstract}

  We present numerical simulations of internal gravity waves (IGW) in
  a star with a convective core and extended radiative envelope.  We
  report on amplitudes, spectra, dissipation and consequent angular
  momentum transport by such waves.  We find that these waves are
  generated efficiently and transport angular momentum on short
  timescales over large distances.  We show that, as in Earth's
  atmosphere, IGW drive equatorial flows which change magnitude and
  direction on short timescales.  These results have profound
  consequences for the observational inferences of massive stars, as
  well as their long term angular momentum evolution. We suggest IGW
  angular momentum transport may explain many observational mysteries,
  such as: the misalignment of hot Jupiters around hot stars, the Be
  class of stars, Ni enrichment anomalies in massive stars and the
  non-synchronous orbits of interacting binaries.

\end{abstract}


\section{Introduction}

Internal gravity waves (IGW) are waves which propagate in stably
stratified regions for which the restoring force is gravity. These
waves are generated by any disturbance in the stratified medium.  For
example, in Earth's atmosphere, IGW are generated by wind flow over
mountain ranges or convective clouds.  In Earth's atmosphere, these
waves are known to drive large scale winds such as the Quasi-Biennial
Oscillation (QBO) of the Earth's equatorial stratosphere
\citep{bald01} and the mesospheric QBO \citep{bald01}.  In our oceans,
waves generated by tidal forces break, leading to turbulent mixing
\citep{mw98}.  In stars, these waves are likely to be generated at the
convective-radiative boundary by both shear stress at the interface
\citep{ktz99} and convective overshoot in the form of plumes
\citep{towns66,mont00,rmac11}.  Therefore, these waves surely exist in
most, if not all, stars and their dynamical relevance will depend on
the configuration of convective and radiative layers, and therefore,
stellar type.

Because of their ability to mix species and transport angular
momentum, IGW have received much attention in stellar astrophysics,
having been proposed to solve a host of long-standing problems.
\cite{press81} suggested IGW could affect the radiative opacity and
hence, the neutrino signature and thus, solve the solar neutrino
problem.  \cite{gls91} suggested IGW could provide the extra mixing
necessary to explain Li depletion in F stars.  \cite{schatz93,mont94}
and \cite{tac05} proposed IGW could cause the Li depletion in the Sun.
\cite{gn89} suggested IGW could explain the efficient tidal
synchronization and circularization of binary stars, and a series of
papers \citep{schatz93,zt97,kq97,tkz02} have suggested that IGW could
be responsible for the uniform rotation of the solar radiative
interior.  More recently, \cite{quats12} have suggested IGW could
cause the enhanced mass loss rates needed to explain some
core-collapse supernovae.

All of the above theories depend on the spectrum and amplitudes of the
IGW driven by convection, which is unconstrained observationally in
stars.  Most of the models develop analytic theories for these
quantities.  In those models, IGW are assumed to be generated either
by turbulent stresses or eddies, as in the studies by \cite{gls91} and
\cite{ktz99}.  IGW are also known to be driven by plumes, as was seen
in early experiments by \cite{towns65} and more recent ones by
\cite{ansuth10}.  The wave spectrum generated by plume incursions was
originally analyzed by \cite{towns66} and extended to stellar
interiors by \cite{monts00}.  Each of these mechanisms, plumes or
eddies, predict slightly different spectra for the waves produced.
However, they all generally show wave amplitudes decreasing with
increasing frequency and decreasing scale, although there are some
cutoff scales and frequencies associated with the eddy frequency and
length scales.

In reality, of course, IGW are likely generated by Reynolds stresses
in the form of eddies ${\it and}$ plume penetration, therefore, the
spectrum that results is likely some combination.  In numerical
simulations by \cite{rmac11} of IGW generated at the base of a
simulated solar convection zone the spectrum was found to be
non-monotonic in scale and a decreasing function of frequency
approximately $\propto \omega^{-3}$ (depending on length scale).  It was
also shown that IGW could not explain the uniform rotation of the
solar radiative interior \cite{rg05,dp08}, nor could they explain the
Li depletion of the Sun \citep{rg06}.  The
varying results between numerical simulations and analytic models with
regard to mixing and angular momentum transport can certainly be
traced to differences in the spectra and wave amplitudes employed
\citep{dp08,tac05,rmac11}.

As mentioned above IGW are known to have profound effects in the
Earth's atmosphere and oceans.  A striking example is that of the QBO
in which IGW propagate upward into the equatorial stratosphere where
they dissipate and thereby, drive a mean flow which oscillates with a
rather regular frequency of approximately 28 months \citep{bald01}.
Given their relevance in our own atmosphere it may appear obvious that
IGW would play a dynamical role in the solar radiative interior.  In
this regard one should note an important difference between IGW in our
own atmosphere and those in solar type stars.  In Earth's atmosphere
IGW are driven by convective clouds or by wind over mountains at
around 10km or $\sim 10^{-4}$ gm cm$^{-3}$, the waves then propagate
upward into a region of ${\it decreasing}$ density and drive the QBO
between 40-60 km or about $\sim 10^{-6}-10^{-7}$ gm cm$^{-3}$.  During
their propagation wave amplitudes increase by $\rho^{-1/2}$ according
to conservation of pseduomomentum \citep{buhlerb}.  Therefore, waves
are likely to break causing substantial angular momentum transport.
Conversely, in solar type stars IGW are generated at the base of the
convection zone and propagate into a region of ${\it increasing}$
density and therefore, wave amplitudes decrease substantially.
Therefore, even if a wave is generated with the convective velocity of
$\approx 10^{3}$ cm s$^{-1}$ at the base of the solar convection zone,
its amplitude will have decayed to 40 cm s$^{-1}$ near the solar core,
even in the absence of dissipation mechanisms.

On the other hand, in massive stars, where IGW have been relatively
unexplored, the physical configuration of wave driver (convection) and
propagation region is similar to the Earth's atmosphere.  For example,
in our models described below, the density between the
convective-radiative interface and the stellar surface decreases by
four orders of magnitude, leading to a wave amplitude amplification of
$\sim$200, in the absence of dissipation.  Given these simple factors
one would rightly expect that IGW in massive stars are more
dynamically and chemically relevant than in solar-type stars.  The
remainder of the paper is organized as follows: in Section 2 we
describe the numerical technique, in Section 3 we briefly discuss
convection and overshoot in these simulations, in Section 4 we discuss
basic properties of IGW and momentum transport by such waves.  In
Section 5 we discuss IGW dynamics in our simulations including,
generation, propagation and dissipation of such waves and the
consequent angular momentum transport.  In Section 6 we discuss the
effects of rotation, viscosity and thermal diffusivity.  We conclude
in Section 7 with a discussion of the consequences of these results to
observations involving massive stars.

\section{Simulating Internal Gravity Waves in Stellar Interiors}

In order to investigate the generation and dissipation of waves in
stellar interiors we solve the full set of hydrodynamic equations,
including rotation, in the anelastic approximation \citep{go69} for an
ideal gas.  The anelastic approximation is valid when flows are
sufficiently sub-sonic and when thermodynamic perturbations are small
compared to the mean (reference state) values, conditions easily
satisfied in stellar interiors.  The radially dependent
reference-state thermodynamic variables (denoted by overbars in
Equations 1-3 below) are taken from the one-dimensional Cambridge
stellar evolutionary code STARS for a 3$M_{\odot}$ star \citep{egg71}
with some minor changes to allow for the calculation of the
Brunt-Vaisala frequency.  These equations are solved in 2D cylindrical
coordinates representing an equatorial slice of the star, extending
from 0.005$R_{\star}$ to 0.98$R_{\star}$.  We solve the following
equations for perturbations to the reference-state:

\begin{equation}
  \nabla \cdot \overline{\rho} \vec{v} = 0 .
\end{equation}

\begin{eqnarray}
  \lefteqn{\dxdy{\vec{v}}{t}+(\vec{v}\cdot\nabla)\vec{v}=-\nabla P - Cg\hat{r} -v_{r}\dxdy{\Omega(r)}{r}\hat{\phi}+}\nonumber\\
  &  &{2(\vec{v}\times\Omega(r))+\overline\nu(\nabla^{2}\vec{v}+\frac{1}{3}\nabla(\nabla\cdot\vec{v}))}
\end{eqnarray}

\begin{eqnarray}
  \lefteqn{\dxdy{T}{t}+(\vec{v}\cdot\nabla){T}=-v_{r}(\dxdy{\overline{T}}{r}-(\gamma-1)\overline{T}h_{\rho})+}\nonumber\\
  &  & {(\gamma-1)Th_{\rho}v_{r}+\gamma\overline{\kappa}[\nabla^{2}T+h_{\rho}\dxdy{T}{r}]+}\nonumber\\
  &  & \gamma\overline{\kappa}[\nabla^{2}\overline{T}+h_{\rho}\dxdy{\overline{T}}{r}] + \frac{\overline{Q}}{c_{v}}
\end{eqnarray}
Equation (1) represents mass conservation in the anelastic
approximation, where $\overline{\rho}$ is the reference state density
and $\vec{v}$ is the velocity in the rotating frame, so the
  velocity in the inertial frame is given by
  $\vec{v}+r\Omega(r)\hat{\phi}$.  Neglecting the time variation of
density essentially filters sound waves.  Equation (2) represents
momentum conservation, where $\Omega(r)$ is the rotation rate, which
could be a function of radius, $\overline{\nu}$ is the viscous
diffusivity, g is gravity, P is the reduced pressure and C represents
the co-density \citep{bra95,rg05}.\footnote{ The reduced pressure is defined as
$P=\frac{p}{\rho}+\psi$ where p is the standard pressure and $\psi$ is
the perturbation of the gravitational potential energy per mass.
Using this formulation of pressure, the ``self-gravity'' term in the
momentum equation, $-\rho\nabla\psi$ is included at no additional
computational cost.  Part of the buoyancy term due to the pressure
perturbation is then absorbed into the pressure gradient term and the
co-density then represents a buoyancy term including both temperature
and pressure components.  This formulation is not as advantageous as
it would be if we were using entropy as our working thermodynamic
variable.  In that case the co-density represents the buoyancy force
due only to entropy perturbations and the pressure component of buoyancy is
completely absorbed into the pressure gradient term.  The interested
reader is referred to the textbook \cite{glatzbook}.}

Equation (3) represents energy
conservation written as a temperature equation, where T is the
temperature perturbation, $\overline{T}$ is the reference state
temperature, $v_{r}$ is the radial velocity, $\gamma$ is the ratio of
specific heats, $\kappa$ is the thermal diffusivity and $h_{\rho}$ is
the inverse density scale height.

The first term on the right hand side of (3) represents the super- or
subadiabaticity of the region and allows us to treat both convective
and radiative regions.  In this formulation of the above equations
that term is set to zero within the convection zone, while the
subadiabaticity in the radiation zone is taken from the stellar model.
In these simulations, the heating that drives convection, which
ultimately comes from nuclear reactions, is contained within the last
term on the RHS of (3), $\overline{Q}$, and takes the form of a
gaussian peaking close to the stellar core and falling to zero at the
convective-radiative interface.  The amplitude of the Gaussian is
determined such that the integrated flux through the system
compensates our enhanced thermal diffusivity.  Because our thermal
diffusivity is too large in these
simulations, and the temperature gradient realistic, the stellar flux is enhanced by the ratio
$\kappa_{sim}/\kappa_{star}$, where $\kappa_{sim}$ is the thermal
diffusivity within the simulation ($5\times10^{11} cm^{2}/s$) and
$\kappa_{star}$ is the stellar diffusivity, which is $\sim 10^{7} cm^{2}/s$.
Therefore the total flux through our simulated star is $\approx 5
\times 10^{4}$ that of an actual star of this temperature, and the
amplitude of $\overline{Q}$ is set to reflect that.  Therefore, we drive our
convection stronger in order to compensate for a stronger dissipation
rate.  Although artificial, some sacrifice of reality must be made in
numerical simulations and we feel this one maintains the best
representation of the dynamics.  We discuss this parametrization in
Section 6.  The boundary conditions are isothermal, stress-free and
impermeable.  The impermeable boundary condition at the top boundary
reflects waves.  However, in a real star the Brunt-Vaisala frequency
would go to zero near the surface, causing internal reflection of the
waves.

Indepedent variables are expanded in a fourier series in azimuth, with
horizontal wavenumber\footnote{Throughout, when we refer to
  wavenumber we will be referring to the horizontal wavenumber.}, $k_{h}$, and decomposed with a finite
difference on a non-uniform grid in the radial dimension.  Horizontal
mean variables are therefore represented by $k_{h}=0$, and where
appropriate, $k_{h}\ne 0$ are referred to as waves.  Explicit
timestepping using the Adams-Bashforth method is applied for the
non-linear terms and implicit timestepping using Crank-Nicolson scheme
is applied for the linear terms.  The various models are run for a
variety of times, with a bare minimum of 10 rotation periods, although
much longer for many models, see Table 1.  The resolution in all
models is 512 horizontal grid points x 1000 vertical grid points, with
400 dedicated to the convection zone and overshoot region.

Although in the future we plan to study models of different stellar
masses, here we fix our reference state stellar evolution model and
investigate the effect of varying rotation rates, studying both
initially uniform and differential rotation rates.  Table~1 lists the
various models run.
\begin{table*}
\begin{minipage}{170mm}

%
  \begin{tabular}{llllllllll}
      \hline
      Model & $\Omega_{i}$ (rad/s) & $\nu $ & $\kappa$ & $\overline{Q}/c_{v}$ & 
      Simulation Time &k spectrum & $\omega$ spectrum & Mean Flow (Peak)\\
      
      \hline
      U1  & 0 & 4 & 0.05 & 3 &1.22 $\times 10^{7}$s  & -1.8/-5.0 & -1.2/-4.8 & SMF (9$\times 10^{-5}$)\\
      U2  & $10^{-6}$  & 4 & 0.05& 3 &1$\times 10^{8}$s (16$\Omega_{p}$) &
      -3.1/-3.5 & -0.9/-2.9 & SMF (8$\times 10^{-5}$)\\
      U3  & $5\times 10^{-6}$  & 4 &0.05 &3 & 2.91$\times 10^{7}$s (23$\Omega_{p}$) &
      -1.9/-4.8 &-1.0/-4.7 & SMF (1$\times 10^{-5}$, SI)\\
      U4  & $10^{-5}$ & 4 &0.05 & 3&5.94$\times 10^{6}$s (9.45$\Omega_{p}$)
      & -1.9/-4.6& -0.8/-3.6
      & SMF (2$\times 10^{-5}$,SI)\\
      U5  & $2\times 10^{-5}$  & 4& 0.05 & 3 &1.56$\times 10^{7}$s (75 $\Omega_{p}$) &
      -1.9/-4.8 & -0.7/-4.2 & SMF (2$\times 10^{-5}$,SI)\\
      U6  & $3\times 10^{-5}$  & 4 & 0.05& 3 &6.25$\times 10^{6}$s
      (30$\Omega_{p}$) & -2.3/-4.6 &-0.5/-3.9  &WMF (4$\times 10^{-6}$,SI)\\
      U7  & $4\times 10^{-5}$  & 4 & 0.05 & 3 &9.26$\times 10^{6}$s
      (59$\Omega_{p}$) & -1.9/-4.7 & -0.6/-4.7 & SMF (8$\times 10^{-5}$)\\
      U8  & $8\times 10^{-5}$  & 4 &0.05 & 3 &1.50$\times 10^{7}$s (191$\Omega_{p}$) &
      -1.9/-4.4 &-0.5/-3.9 & SMF (1.1$\times 10^{-4}$)\\
      U9  & $10^{-4}$  & 4 & 0.05 & 3 &1.35$\times 10^{7}$s (215$\Omega_{p}$) &
      -2.8/-5.7 & -1.3/-2.4 & NMF\\
      D10 & $3/2\times 10^{-6}$ & 4& 0.05 & 3&1.6$\times
      10^{7}$s (5.0$\Omega_{p}$) & -2.5/-3.8 & -1.0/-3.4 & SMF
      (2$\times 10^{-5}$)\\
      D11 & $3/2\times 10^{-5}$  & 4 &0.05 & 3 &3.8$\times
      10^{7}$s (121$\Omega_{p}$) & -3.1/-3.8 & -1.0/-3.3 & SMF (1.1$\times 10^{-4}$)\\
      D12 & $2.5/2\times 10^{-5}$ & 4 &0.05 & 3 &1.22$\times
      10^{7}$s (39$\Omega_{p}$) & -2.1/-3.8 & -0.8/-3.2 & SMF (2$\times 10^{-5}$,SI)\\
      D13 & $2.1/2\times 10^{-5}$  & 4 &0.05 & 3 &4.0$\times 10^{6}$s
      (13$\Omega_{p}$) & -3.0/-3.8 & -1.0/-3.0 &WMF (1$\times 10^{-6}$,SI)\\
      MU1-1 & 0 & 4 & 0.05 & 1 &4$\times$10$^{7}$s  &-1.6/-5.8 & -0.9/-4.4 &NMF\\
      MU1-2 & 0 & 4 & 0.016 & 1 &3.9 $\times 10^{7}$s  &-1.7/-4.1 & -0.9/-3.9 &NMF\\
      MU1-3 & 0 & 4 & 5.0 & 3 &2.1 $\times 10^{6}$s  &-5.4/-36.0 & -1.3/-5.1&NMF \\
      MU1-4 & 0 & 40 & 0.05 & 3 &2.1 $\times$ 10$^{7}$s  &-1.6/-5.0 & -0.4/-4.1 &NMF\\
      MU1-5 & 0 & 8 & 0.05 & 3 &7 $\times$ 10$^{7}$s  &-2.2/-4.0 & -0.9/-3.7 &NMF\\
      MU1-6 & 0 & 2 & 0.05 & 1.5 &5.5 $\times$10$^{6}$s  &-1.8/-4.9 & -1.1/-4.2 &SMF
      
      (8$\times 10^{-6}$, SI) \\
      \hline
    \end{tabular}

    \normalsize
    \caption{Model parameters. ``U'' models represent {\it initially}
      uniformly rotating models, ``D'' denotes initially
      differentially rotating models and ``MU1'' models represent
      modified versions of model U1.  Rotation rates are give in
      rad/s.  Initially differentially rotating models have rotation
      rates which are constant within the convection zone and
      radiation zone with the listed values matched at the
      interface with an arctan function of width 0.02$R_{\ast}$.
      Diffusivities, $\nu$ and $\kappa$, are given in units of
      $10^{13} cm^{2} s^{-1}$.  $\overline{Q}/c_{v}$ is given in units
      of $10^{4} K s^{-1}$. k spectrum and $\omega$ spectrum
      values represent exponents for the power law fits for low/high
      wavenumbers and frequencies, respectively.  Mean Flow states are
      referred to as ``SMF'' if a strong mean flow develops (a flow at
      least as strong as the initial background rotation) , ``WMF'' if
      a weak mean flow develops and ``NMF'' if no mean flow develops.
      The values listed are the peak values attained during the
      simulation time in units of s$^{-1}$ and ``SI'' means the mean flow was still
      increasing in amplitude when the simulation was terminated.}
\end{minipage}
\end{table*}

\section{Convection and Penetration}
The convection in these simulations is dominated by convective plumes
which emerge from the core and generally span the depth of the
convection zone.  Once the plumes impinge on the overlying stable
region they become unstable as they are decelerated by the rapid
variation in the Brunt-Vaisala frequency.  As the plumes become
unstable they break up into small eddies which propagate laterally
along the convective-radiative interface.  This behavior can be seen
in the time snapshots shown in Figure~\ref{fig:fig1-longp} , although
it is best seen in animation, which can be found
www.solarphysicist.com.  The destabilization of the plume and
subsequent eddy production leads to substantial mixing and convective
overshoot in the model.

Convective overshoot in massive stars has received much attention both
theoretically and observationally \citep{rox65,rox92,woo01}.  Such
overshoot can lead to chemical mixing which could provide fuel from
the radiative envelope into the convective core, thus extending the
lifetime of the star.  Although there are several ways to define the
overshoot depth, here we define it as the first place the kinetic
energy flux changes sign outside the convection zone.  In
Figure~\ref{fig:pen-rot} we show the overshoot depth as a function of
rotation rate (and time) for a subset of the uniformly rotating
models.  There we see that the overshoot depth varies substantially
between $\approx 0.1 -0.5$H$_{p}$, where H$_{p}$ is a pressure scale
height (see figure caption for symbol distinction).

Although there are clear exceptions, in general, the overshoot depth
decreases for increasing rotation rate and increases in time (later
times denoted by diamonds).  This occurs because radial motions are
deflected azimuthally by the Coriolis force, thus limiting their
ability to overshoot the convective-radiative interface.  Indeed, we
find that radial velocities at the interface are nearly 40\% smaller
in our highest rotating models, relative to our non-rotating models.
This rotation rate dependence was also found in \cite{brumm02}.  We
also note that the while the penetration depths increase in time, many
of our models show saturation at later times.

Although more work needs to be done for various stellar types, our
results indicate that there is a wide range of penetration depths that
can be achieved in a given model, that the penetration depth can
evolve in time and that the rotation rate can have an order 2 effect.
It is interesting to note that the effectiveness of mixing by
Eddington-Sweet type circulations (at a given age) increases with
increasing rotation rates, whereas mixing by convective overshoot
decreases with increasing rotation rate.  These two effects may create
a ``sweet spot'' in rotation space and age, where mixing beyond the
convection zone is optimized.

In the above we measured convective overshoot, that is, how far
convective motions could extend into the radiative zone, defined by
the initial temperature stratification.  As discussed in \cite{za91}
continual convective overshoot could transfer enough heat to the
region to bring about a change in the mean thermal stratification,
what is often referred to as ``penetration''.  We show the mean
temperature stratification as a function of radius for several models
in Figure~\ref{fig:stratnew} (see Figure caption for details).  We see
that models with lower rotation and/or differential rotation have, on
average, more penetration.

In these models we see substantial convective overshoot and
penetration, both of which depend on the rotation rate, with faster
rotators showing lower penetration and overshoot.  These results are
consistent with observational constraints that place convective
overshoot near 0.2-0.3H$_{p}$ \citep{mmm93}, although our values may
be on the high side due to slightly larger convective velocities (see
Section 2 and Section 6).  We note that because a shear flow is set up
between the convective and radiative regions, and because the
Brunt-Vaisala frequency is low near the interface, the overshoot
region is unstable to shear instability.  This would also cause
substantial mixing in the region, but such mixing may or may not be
relevant in the presence of convective overshoot and penetration.

\section{Internal Gravity Wave Dynamics}
\subsection{Fundamentals of IGW and Wave-Driven Mean Flows}
The simulations presented here are of inertial gravity waves, which
have a simplified dispersion relation:
\begin{equation}
  \omega^{2}=\frac{k_{h}^{2}N^{2}}{k_{h}^{2}+k_{v}^{2}}+\frac{f^{2}k_{v}^{2}}{k_{h}^{2}+k_{v}^{2}}
\end{equation}
where $\omega$ is the frequency of the wave\footnote{Note that this
  dispersion relation is for an f-plane model, which is not the
  geometry of our simulation but it yields effectively the same
  dispersion relation assuming f=2$\Omega$.}, f is $2\Omega$,
$k_{h},k_{v}$ are the horizontal and vertical wavelength,
respectively, and N is the Brunt-Vaisala frequency.  In the absence of
rotation, this dispersion relation is often written as
\begin{equation}
  \omega=\pm N\sin\phi
\end{equation}
where $\phi$ is the angle between the horizontal plane and lines of
constant phase.  Therefore, shallow angles imply a low frequency,
while steep angles imply a high frequency.  Another simplification of
the dispersion relation can be found in the limit of $k_{h} < k_{v}$,
the dispersion relation then reduces to $\omega/N \sim k_{h}/k_{v}$.
For pure IGW, waves can only propagate when $\omega < N$.  When
rotation is considered waves can only propagate when $f < \omega < N$.
Waves with frequencies approaching N are internally reflected, whereas
waves with frequencies lower than the rotation frequency encounter
critical layers, where most of the wave energy is
dissipated.\footnote{A critical layer is defined as the position where
  the local angular velocity is equal to the horizontal phase speed of
  the wave.  In the ray-tracing description of IGW the vertical
  wavenumber goes to infinity at a critical layer, and, therefore, the
  time to approach the critical layer goes to infinity.  This means
  any small dissipation leads to complete, or nearly complete,
  absorption of the wave at the critical layer.}

From the dispersion relation one can show that the phase velocity in
the absence of rotation is
\begin{equation}
  c_{p}=\omega/{\bf k}=\omega\left(\frac{k_h}{k_h^{2}+k_{v}^{2}},\frac{k_{v}}{k_h^{2}+k_v^{2}}\right)
\end{equation}
and the group velocity is
\begin{equation}
  c_{g}=\dxdy{\omega}{\bf k}=\left(\frac{N
      k_{v}^{2}}{(k_{h}^{2}+k_{v}^{2})^{\frac{3}{2}}},\frac{-N k_{v} k_{h}}{(k_{h}^{2}+k_{v}^{2})^{\frac{3}{2}}}\right)
\end{equation}
one can then see that the phase and group velocities are
perpendicular, while the vertical components are in opposite
directions (note that vectors are represented as (horizontal
component, vertical component)).  Therefore, if the group velocity is moving outward the
phase velocity is moving inward and the fluid motions are along lines
of constant phase.  As a wave propagates outward from the convection
zone the Brunt-Vaisala frequency is increasing, therefore, the ratio
$\omega/N$ is decreasing, consequently the fluid motions and phase
path are refracted, causing the wave motion to be a spiral path
outward from the convection zone.  It is worth noting that while
stratification forces fluid motions to be more horizontal, rotation
forces fluid motions to be more vertical, so that these two forces
(buoyancy and Coriolis) have opposite effects on the fluid motions.

How these waves transport angular momentum has been a topic of intense
research in the atmospheric science community.  Here we review some of
the basics of wave-driven mean flow (the interested reader is referred
to standard atmospheric science textbooks, such as
\cite{pedloskyb,holtonb,lindzenb}).  When a single propagating wave is
attenuated it transfers angular momentum to the mean flow
\citep{ep60}.  The forcing of the mean flow by waves depends on the
wave attenuation, which depends on the frequency of the waves relative
to the mean flow, which in turn, depends on the mean flow.  Therefore,
wave-mean flow dynamics is highly nonlinear.

One can appreciate how angular momentum is transferred from waves to
the mean flow by considering the longitudinally-averaged horizontal
component of the momentum equation \citep{holtonb,lindzenb}:
\begin{equation}
  \dxdy{\overline{
      U}}{t}+\frac{1}{r\overline{\rho}}\dxdy {~r \overline{\rho
      v_{\phi}
      v_{r}}}{r}=\frac{\nu}{r\rho}\frac{\partial}{\partial{r}}\left(r \overline\rho\dxdy{\overline{U}}{r}\right)
\end{equation}
where $v_{\phi}$ is the azimuthal velocity, and $v_{r}$ is the radial
velocity and overlines denote a horizontal average.  This equation
shows that the mean zonal flow, $\overline{U}$, is driven by the
divergence of the horizontally-averaged Reynolds stress and is
decelerated by viscous dissipation.  

The mean zonal flow maintained by internal gravity waves is typically
depth dependent, i.e., a ${\it shear}$ flow.  This "anti-diffusive"
nature of gravity waves can be explained heuristically.  Imagine a
prograde wave ($k_{h}>$0) and a retrograde wave ($k_{h}<$0) excited at
the same radius.  If the medium in which the waves propagate is
differentially rotating, these waves will be doppler shifted by:
\begin{equation}
  \omega(r)=\omega_{gen}+k_{h}[\Omega_{gen}-\Omega(r)] .
\end{equation}
Therefore, if the angular velocity, $\Omega(r)$, of the medium
increases in the direction of propagation, the prograde wave is
Doppler shifted to lower frequency, whereas the retrograde wave is
shifted to higher frequency.  These frequency shifts are relative to
the frequency at which the waves were generated, $\omega_{gen}$, where
the angular velocity is $\Omega_{gen}$.

Considering the simplest dissipation mechanism, radiative damping, the
dissipation length goes approximately as $d_{L}\approx v_{gh}/ \gamma$
where $v_{gh}$ is the horizontal group velocity\footnote{Note, the
  vertical group velocity is often used here, but the wave propagation
  is predominantly horizontal, therefore, it is more realistic to use
  the horizontal group velocity \citep{rmac11}.  Furthermore, we find
  both here, and in previous work, that an  $\omega^{3}$ dependence which
  arises from considering the horizontal group velocity, fits the data
from numerical simulations better than a $\omega^{4}$ dependence.} and $\gamma$ is the
dissipation rate, $ \approx \kappa k_{v}^{2}$.  Using Equation (7) for
the horizontal group velocity, we arrive at a dissipation length which
depends on frequency and wavenumber like:
\begin{equation}
  d_{L}\approx \frac{\omega^{3}}{\kappa N^{2}k_{h}^{3}}
\end{equation}
Therefore, waves which are dopper shifted to higher frequency
(retrograde waves in the above example) will have a longer dissipation
length than those shifted to lower frequency and therefore, will
propagate a longer distance before being dissipated.  Alternatively,
the higher frequency wave will have a higher amplitude after traveling
a set distance relative to the lower frequency wave.
  
A prograde wave transports positive angular momentum, whereas a
retrograde wave transports negative angular momentum.  Therefore,
where a prograde wave is dissipated the mean zonal flow is accelerated
and where a retrograde wave is dissipated the mean flow is
decelerated.  In this way two waves excited at a given radius with the
same frequency and wavenumber but spiralling in opposite directions
outward can lead to a radial shear flow.  This process is
self-reinforcing and so the shear will grow until viscosity becomes
important.  \cite{plumb77} showed that two waves are unstable and will
produce a shear, even in the absence of rotation or an initial shear.
This was also demonstrated in the remarkable experiment by
\cite{pm78}.

\section{IGW Dynamics in Massive Star Simulations}
\subsection{Overview}
\begin{figure}
  \centering
  \includegraphics[width=7in]{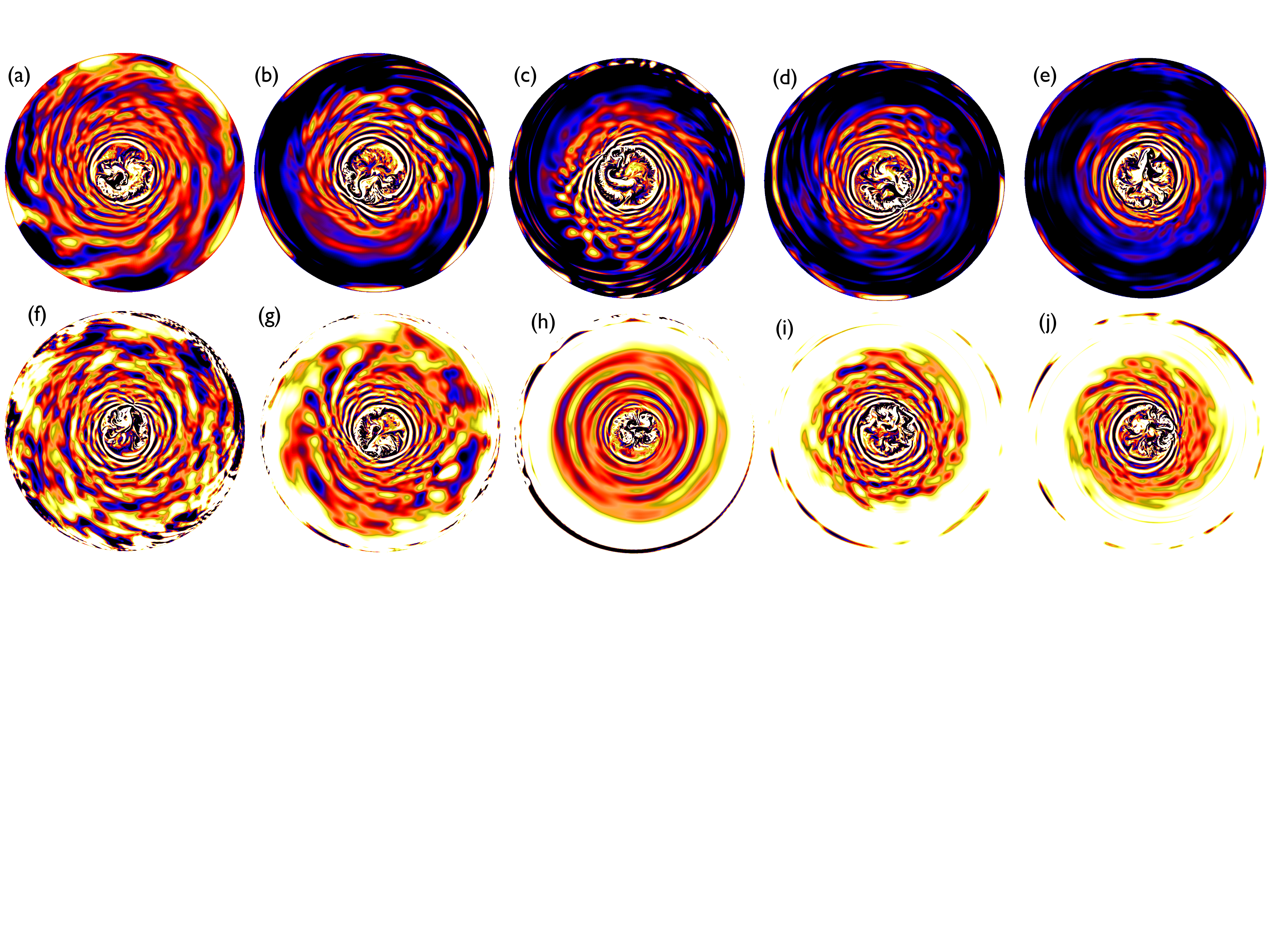}
  \caption{Time snapshots of vorticity for Model U2 (corresponding to
    times 6.6 (a), 7.1 (b), 7.3 (c), 7.5 (d) and 8.3 (e) rotation periods) and Model U8
    (corresponding to 13 (f), 38 (g), 75 (h), 80 (i) and 88 (j)
    rotation periods).  Black represents negative vorticity, while white
    represents positive vorticity.}
  \label{fig:fig1-longp}
\end{figure}

\begin{figure}
  \centering
  \includegraphics[width=3.5in]{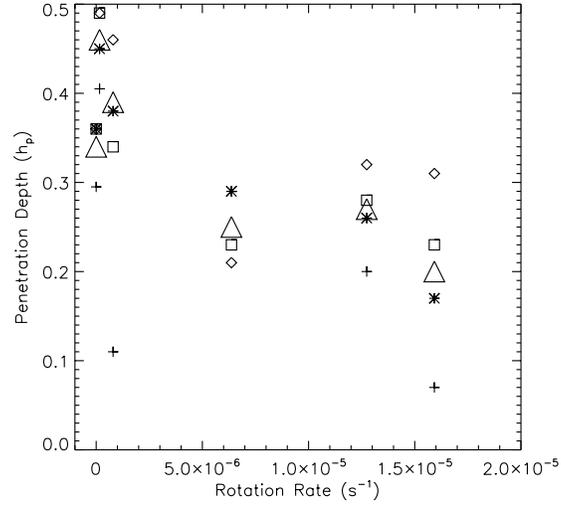}
  \caption{Overshoot depth in terms of pressure scale heights,
    calculated as the first zero of the kinetic energy flux averaged
    in longitude at various times for various rotation rates.  Pluses,
    asterisks, squares and diamonds represent various times during the
    simulations from early to late, respectively (note that they are
    not necessarily the same times).  Large triangles represent
    averages of these various times.}
  \label{fig:pen-rot}
\end{figure}

\begin{figure}
  \centering
  \includegraphics[width=3.5in]{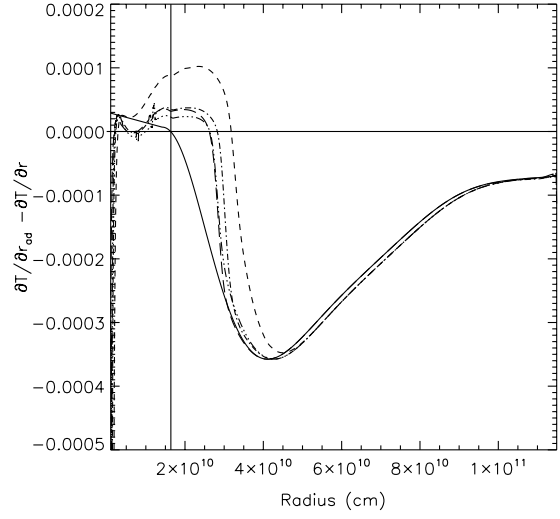}
  \caption{Super(Sub)-adiabaticity as a function of radius for models
    U2 (dashed line), U1 (dot-dashed line), U9
    (triple-dot dashed) and U3 (long dashed), averaged over ~$10^{6}$s.  The solid line shows
    the initial stratification for all models.  The solid vertical
    line marks the convective-radiative boundary, while the horizontal
  vertical line marks the boundary between super and subadiabaticity.}
  \label{fig:stratnew}
\end{figure}

Figure~\ref{fig:fig1-longp} shows the evolution of Models U2 and U8.
In the following paragraphs we will briefly describe the wave-mean
flow interaction that leads to the behavior shown in those figures
with more details given in the subsequent sections (again, the
interested reader is referred to the above texts or \cite{buhlerb},
which discuss wave-mean flow interaction in great detail).  Waves are
generated at the convective-radiative interface and propagate into the
radiative envelope toward the surface.  Waves are generated both by
direct plume incursion and by eddies propagating laterally along the
interface.  These processes produce a rather broad spectrum in
frequency and wavenumber.  The waves then propagate nearly
horizontally in a spiral path outward due to the small ratio of
$\omega/N$ and the variability of N with radius (see, for example,
Figure~\ref{fig:fig1-longp} c,d,j).

As waves propagate toward the surface their amplitudes are affected
primarily by two effects: radiative diffusion which acts to dissipate
waves and density stratification, which acts to amplify waves.  The
combination of these two effects implies that the waves that make it
to the surface of the star with the highest amplitude are primarily
high frequency, large scale waves.  These waves make it to the surface
with high enough amplitude that they are subject to wave breaking and
consequent angular momentum transfer.  This angular momentum transfer
drives a mean flow (seen as negative vorticity, black, in
Figure~\ref{fig:fig1-longp}b-e and positive vorticity, white, in
Figure~\ref{fig:fig1-longp}g-j).

In an initially uniformly rotating model the sense of the initial mean
flow that develops (prograde or retrograde) is random \citep{plumb77}.
In the models shown in Figure~\ref{fig:fig1-longp}, U2 shows a mean
${\it retrograde}$ flow at the surface, while U8 shows a mean ${\it
  prograde}$ flow at the surface.  As waves propagate into a
differentially rotating medium they are doppler shifted as described
in Equation (9).

Considering the model U8 (bottom panel in Figure~\ref{fig:fig1-longp},
bottom panel in Figure~\ref{fig:angvel-nv8-nv14}).  Once the
prograde mean flow develops, prograde waves propagating into that flow
are shifted to lower frequencies and therefore, dissipated in shorter
distances, while retrograde waves are shifted to higher frequencies
and thus dissipated less strongly\footnote{Note that the waves are
  dissipated continuously and the dissipation length is just defined
  in the normal sense of when the amplitude has decreased by a factor
  of e.}.  The prograde waves deposit positive angular momentum thus
causing the prograde shear flow to ${\it grow}$ in time.  As the shear
grows, the differential rotation increases, and prograde waves are
shifted to lower and lower frequencies, and therefore, dissipated in
shorter and shorter distances, causing the mean flow to migrate toward
the source of the waves in time (note the difference in the extent of
the mean flow between Figure~\ref{fig:fig1-longp}b and d, or g and j).
Therefore, this process is self-reinforcing, until viscosity becomes
relevant.

As the entire region is spun up, angular momentum must be conserved
and therefore, the convecting core slows down.  This configuration, in
which the angular velocity is increasing outwards allows retrograde
waves to now reach the surface with higher amplitude than their
prograde counterparts.  This leads to transport of predominantly ${\it
  negative}$ angular momentum and the consequent slowing of the outer
regions, eventually leading to a reversal of the mean flow at the
surface (see Figure~\ref{fig:angvel-nv8-nv14}).\footnote{We have
  not been able to model a full reversal at this time due to an
  instability which develops in the flow.  This instability is
  currently under investigation.}

The wave-mean flow interaction described above has been elucidated
well in many text books, and was first described by \cite{hl72} to
explain the Quasi-Biennial Oscillation of the Earth's equatorial
stratosphere.  The process was then demonstrated in the remarkable
experiment by \cite{pm78}.  What we observe in our simulations is not
a new physical process, it is only the application of this physical
process to massive stars.  

\subsection{Wave Generation}

The generation of waves at convective-radiative interfaces has
received much attention in stellar physics as it is the amplitude and
spectra of waves generated that dictates the details of angular
momentum transport and mixing by these waves.  The work of
\cite{press81} and \cite{gls91} use physical arguments matching the
ram pressure of the convective fluctuations to the wave pressure,
combined with a Kolmogorov spectrum of turbulence to arrive at a wave
spectra.  Work by \cite{ktz99,mont00,lecq12} solve an inhomogeneous
wave equation with a source term to represent the wave driving.  The
source term in those models takes the form of Reynolds stresses
provided by convective eddies, as in \cite{ktz99,lecq12} or that of a
plume model as in \cite{towns66,mont00}.  Not surprisingly, the
various models lead to different spectra.  For example, the model of
\cite{ktz99,lecq12} in which the source term is that of convective
eddies leads to spectra of the form:
\begin{equation}
  E(\omega,k_{h}) \propto \omega^{-a} k_{h}^{b} (1+Bk_{h})exp[-A k_{h}^{2} \omega^{-3}]
\end{equation}
where $a \approx 13/2$ and b=2.  This models shows that
$E(\omega,k_{h}$) is a decreasing function of frequency and a function
of scale which increases at low wavenumber to some typical length
scale and then decreases at higher wavenumber.  The plume model of
\cite{towns66,mont00} leads to a wave energy spectrum of the form:
\begin{equation}
  E(\omega,k_{h}) \propto \exp(-\frac{1}{2} (\tau^{2} \omega^{2}+b^{2} k_{h}^{2}))
\end{equation}
in which $\tau$ is the characteristic timescale of plume intrusion and
$b$ is a characteristic spatial scale.  In this model
$E(\omega,k_{h})$ decreases exponentially with frequency and scale.

Besides the obvious differences in these spectra there is also a more
qualitative difference in the frequencies and scales thought to be
associated with the different processes of plume excitation and
convective eddie excitation.  In the plume theory the frequencies are
associated with the timescale of plume intrusion.  Given the steep
gradient of the Brunt-Vaisala frequency, the plume is halted shortly
after entering the radiative region and therefore, plumes can drive
high frequency waves.  The typical wavelength is associated with the
horizontal extent of the plume.  Given the thin-ness of the plume this
would indicate small horizontal scales (although see discussion
below).  On the other hand, convective eddies that span some fraction
of a pressure scale height and have slower turnover times would drive
larger scale waves and shorter frequencies.  Therefore, in the
theories described above, convective eddies are responsible for low
frequency, large (horizontal) wavelength waves, while plumes are
associated with high frequency, short wavelength waves.

Turning to our simulations we can get a qualitiative picture of the
wave generation process (again, although images are shown here, the
clearest picture of the physical processes can be seen in animation at
www.solarphysicist.com).  The convection zone is dominated by plumes
which span the extent of the convective core, often impeding on the
overlying convection zone.  As the plume intrudes into the stable
region it is rapidly decelerated by buoyancy causing the plume to
break up, generating laterally propagating vortices at the interface
which are a fraction of the convection zone depth.  Therefore, both
plumes and eddies contribute to the spectrum of wave energy.

\begin{figure}
  \centering
  \includegraphics[width=4in]{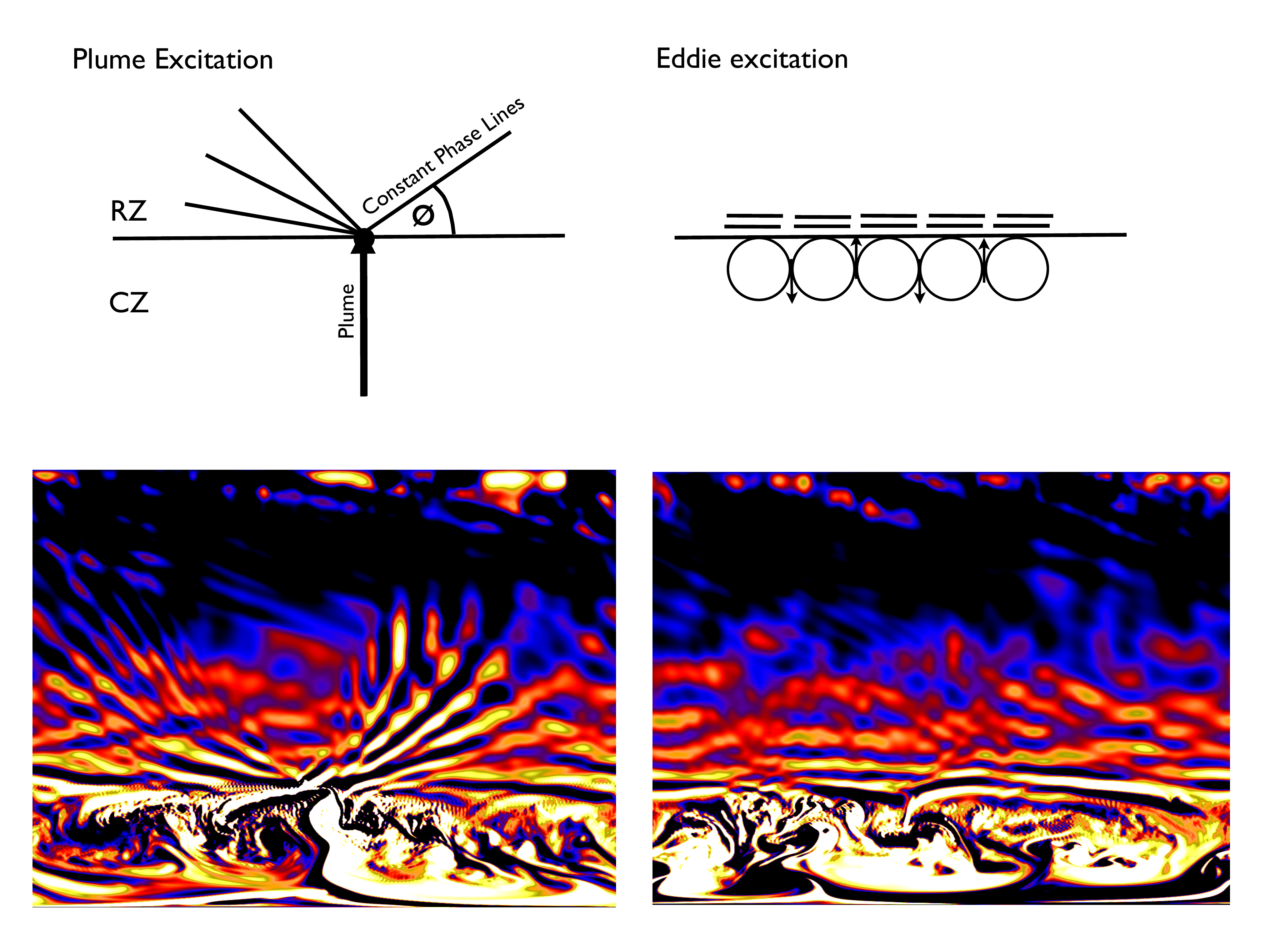}
  \caption{Top panels show diagrams of plume excitation of IGW (left)
    and eddie excitation of IGW (right).  The angle $\phi$ indicates
    the angle between constant phase lines and the horizontal plane,
    when $\phi$ is large, wave frequencies are large, and vice versa.
    Lower panels show snapshots of vorticity within our simulation,
    U1.  One can easily see high frequencies associated with plume
    incursions (left bottom) and low frequencies associated with
    eddies (right bottom).  Note that the time snapshots are from our
    cylindrical model mapped onto a cartesian grid so the upper radial
    levels are compressed and the image slightly distorted.}
  \label{fig:plume-eddie-diagram}
\end{figure}

Wave generation by each of these processes can be understood
qualitatively.  As plumes intrude on the overlying convection zone
they drive waves with ${\it vertical}$ wavelengths similar to the
plume intrusion depth (which is small, see
Figure~\ref{fig:plume-eddie-diagram}), this gives rise to large
horizontal scale waves, since $\omega/N\sim \lambda_{v}/\lambda_{h}$
where $\lambda_{v,h}$ are the vertical and horizontal length scales.
In animation, and figures, plume intrusions can clearly be associated
with large horizontal scale waves.  On the other hand, shorter
wavelength waves are seen to be driven as the plume breaks up into
laterally propagting small scale eddies.  Qualitatively, this means
plumes are associated with large scales, while eddies are associated
with small scales.

The frequency of waves generated by the plume incursion can be
seen in Figure~\ref{fig:plume-eddie-diagram}.  When the plume intrudes, there are a
variety angles of phase lines with respect to the horizontal that are
generated, indicating plumes generate a variety of frequencies (see
Figure~\ref{fig:plume-eddie-diagram} left panels).  In the dispersion
relation for IGW expressed in Equation (5), the angle $\phi$ is the
angle between the horizontal and lines of constant phase.  Therefore,
shallow angles indicate low frequencies, while steep angles indicate
higher frequencies.  On the other hand, when there are no plume
incursions waves are driven by stresses provided by convective eddies
(see Figure~\ref{fig:plume-eddie-diagram} right panels).

\begin{figure}
  \centering
  \includegraphics[width=4in]{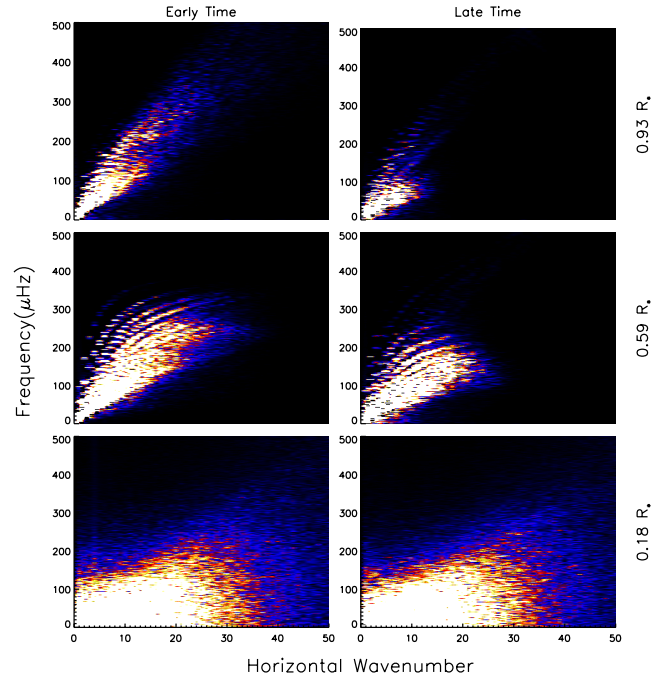}
  \caption{Power spectra, E($k_{h},\omega$),for U2 at three heights,
    just outside the convection zone at $0.18R_{\star}$, midway
    through the radiation zone $0.59R_{\star}$ and near the top of the
    radiation zone $0.93R_{\star}$, at early times (defined as before
    a strong mean flow has developed at the surface, left panels) and
    later times (defined as after a strong mean flow has
      developed at the surface, right panels).  The spectrum at generation (bottom
    panels) is broad up to $k_{h}\approx 30$, ridges associated with
    standing g-modes are clearly seen at mid-radiation zone.  Wave
    energy at the highest frequencies and smallest scales is
    dissipated at later times.}
  \label{fig:powspec_3heights_nv8}
\end{figure}

We can evaluate the wave spectrum more quantitatively by looking at
the wave energy as a function of frequency and wavenumber.  Because
our numerical code is spectrally decomposed in the horizontal
dimension this can be done easily by taking a fourier transform of a
time series of data at a particular radius.  The results of this
procedure are shown in Figure~\ref{fig:powspec_3heights_nv8}, at
different radii at both early and late times.  In
Figure~\ref{fig:powspec_3heights_nv8} we see that at the
convective-radiative interface, waves are generated with a broad
spectrum down to horizontal wavelengths of $\sim 4.6\times 10^{9}$cm
($k_{h} \sim 30$) and up to frequencies of $\sim 200\mu Hz$.  Moving
into the bulk of the radiative envelope we see that lower frequency
waves have been dissipated and we see ridges in the power spectrum,
indicative of standing g-modes.  Finally, at the surface we see only
fairly high frequency waves.

\begin{figure}
  \includegraphics[width=9in]{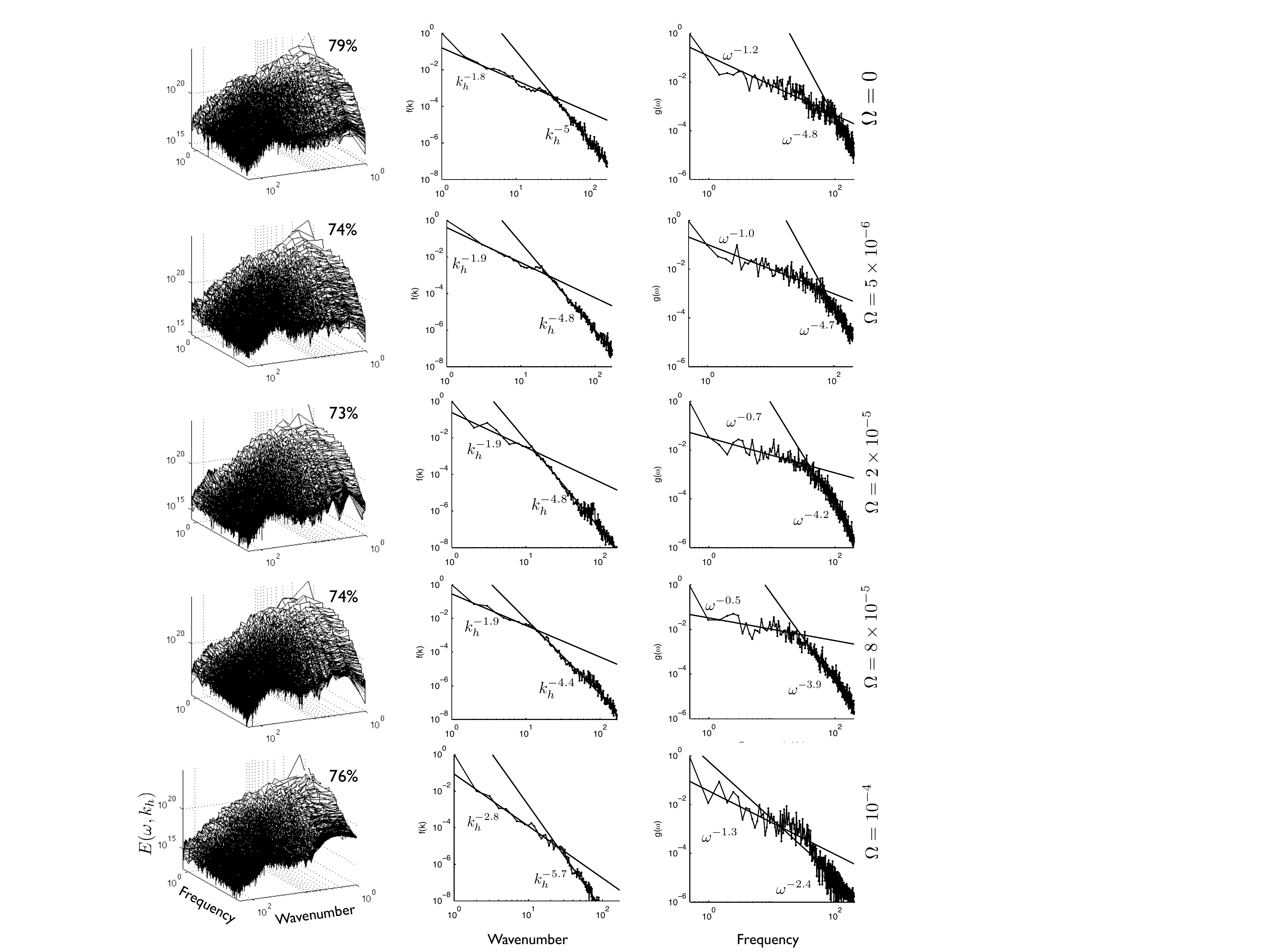}
  \caption{Left column: Wave energy as a function of frequency and
    horizontal wavenumber 0.5 $H_{p}$ outside the convection zone.  Frequency
    resolution is 0.5 $\mu$Hz.  Middle Column: Normalized fit in
    wavenumber to the SVD of wave energy.  Right Column: Normalized
    fit in frequency to the SVD of wave energy.  The percentage of
    energy fit by a separable function is listed in the left hand
    column, while the best fit power law exponents are labeled in each
    panel.  Rotation rate increases downard.}
  \label{fig:comp-omkspec}
\end{figure}

In this section we address the spectrum and amplitudes of wave {\it
  generation} at the convective-radiative interface and leave the
issue of their propagation and attenuation with radius for the next
sections.  Therefore we concentrate on the energy spectrum shown in
the bottom panels of Figure~\ref{fig:powspec_3heights_nv8}.  In
Figure~\ref{fig:comp-omkspec} we show the wave\footnote{The word ``wave''
  here is used loosely and only means horizontal wavenumber not equal
  to zero, these "motions" do not necessarily obey a simple, linear dispersion
  relation.} spectra as a function of frequency and wavenumber at 0.5
pressure scale heights above the convective-radiative interface for
several of our models with varying rotation rate (see Figure and
caption for details).  One can see that waves are excited at all
frequencies and wave numbers but the amplitudes fall off rapidly at
high wavenumber (small scales) and high frequencies as predicted in
both the theories described above.

By inspection one might expect that the energy shown in the left
panels of Figure~\ref{fig:comp-omkspec} is a separable function in
frequency and wavenumber and we indeed confirm that this is the case
using a singular value decomposition (SVD) of the wave energy.  We
find that a separable function can explain of order 70-80\% of the
power seen in Figure~\ref{fig:comp-omkspec}, as denoted by the
percentages shown in that Figure\footnote{Moving further into the
  radiative interior the function becomes less separable, as expected
  given that the dispersion relation strongly links frequencies and
  wavenumbers.}.  Assuming the wave energy can be written as the
separated function
\begin{equation}
  E(\omega,k_{h})=f(\omega)g(k_{h})
\end{equation}
where $f(\omega)$ and $g(k_{h})$ come from the SVD, we find that the
best fit solutions to the data are power laws, as shown in the middle
and right panels of Figure~\ref{fig:comp-omkspec}, respectively.  We
generally find that the wave energy can be approximated well as
$E(\omega,k_{h})\propto \omega^{-m} k_{h}^{-n}$, with both wavenumber
and frequency spectra fit best by two separate power laws.  We
find that at low frequencies $m \sim 0.87\pm 0.24$ and at high
frequencies $m$ is $\sim 3.7\pm 0.72$.  Similarly, at low wavenumbers
$n \sim 2.3 \pm 0.5$ and at high wave numbers (smaller scales) $n
\sim 4.4 \pm 0 .6$.  The mean and standard deviations are calculated
from all our models in which the heat flux is the same (Models U1-U9
and D10-D14).  Within a given model there is also variation in power
law spectra in time.  When sampled at 12 non-overlapping times, model
D11 shows $m\sim 0.89 \pm 0.14$ at low frequencies and $m\sim 3.25 \pm
0.17$ at high frequencies, $n\sim 2.7 \pm 0.26$ at large scales and
$n\sim 3.8 \pm 0.06$ at small scales.  Where the mean and standard deviations quoted refers
to the variation in the spectra over these 12 times.  From these values we deduce
that time variation of a spectra is stronger than variations due to
rotation.  However, we note that the trend in power laws in both
frequency and wavenumber is always seen.  We expect that these two
power laws are associated with the two different processes of plume
and eddie excitation, at least broadly.

\begin{figure}
  \includegraphics[width=4in]{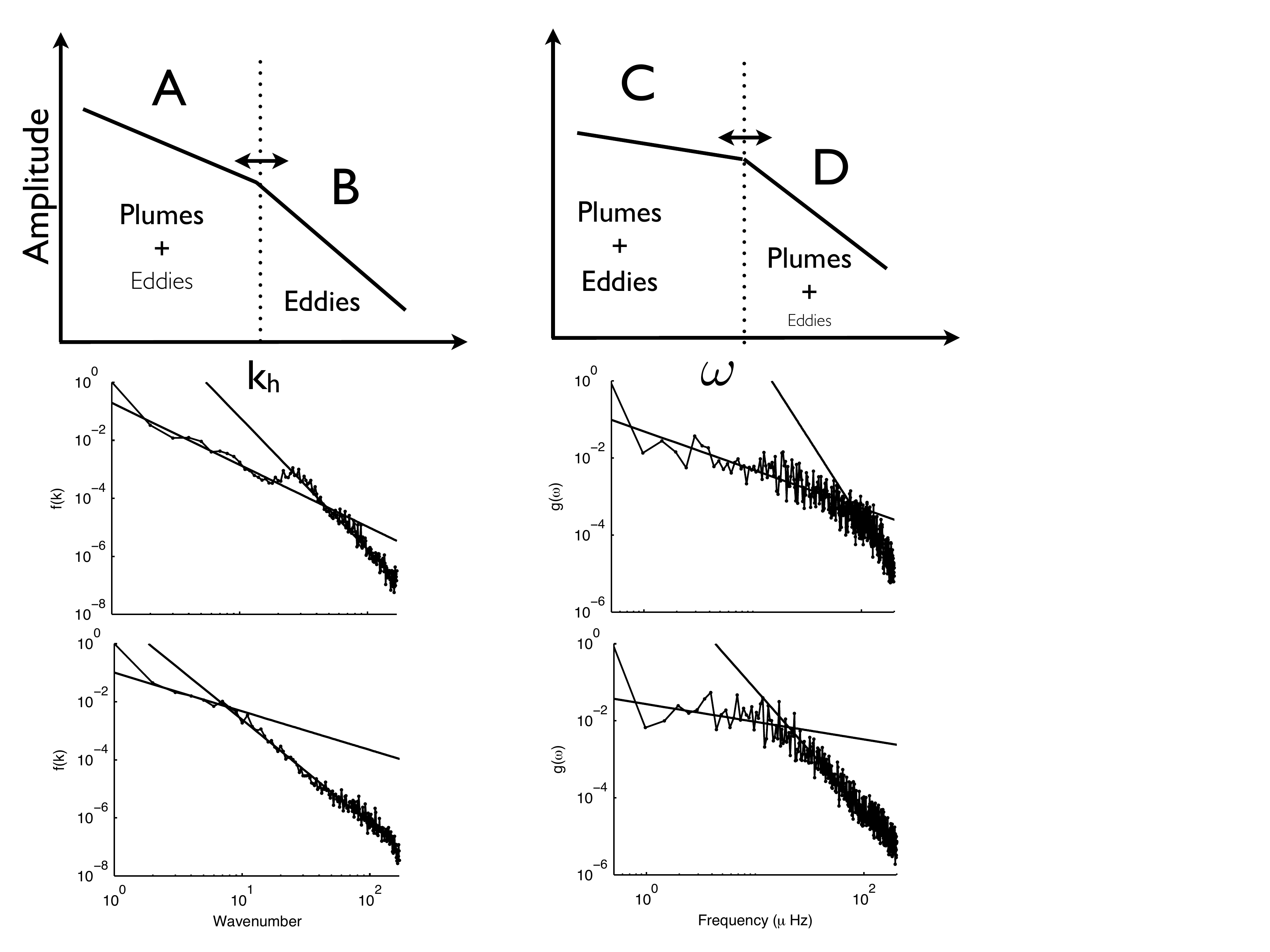}
  \caption{Diagram of wave spectrum generated by plumes and eddies.
    Size and weight of font indicates qualitatively the relative
    importance of different mechanisms in different physical regimes.}
  \label{fig:plume-eddie-specdiagram}
\end{figure}

Figure~\ref{fig:plume-eddie-specdiagram} shows a diagram of which
processes we think dominate which region of the spectrum seen in
Figure~\ref{fig:comp-omkspec}.  Large horizontal length scales, Region
A, are dominated by plumes, with eddies contributing to a lesser
extent by the coherent sum of many waves.  As described above, the
vertical wavelength of a plume is related to the plume intrusion
depth.  This length scale is short, therefore, with the approximate
dispersion relation given by $\omega/N\approx \lambda_{v}/\lambda_{h}$,
$\lambda_{h}$ is therefore, larger than the vertical length scale by
the amount $N/\omega$.  If the vertical wavelength is the penetration
depth, $\Delta_{p}$, then the horizontal wavelength is given by
$\lambda_{h}\approx \Delta_{p}N/\omega$.  Therefore, the {\it minimum}
horizontal wavelength is $\Delta_{p}$, hence the {\it maximum}
horizontal wavenumber associated with plumes is approximately the
wavenumber associated with the penetration depth.  From Section 3, the
penetration depth varies from $\sim 0.1-0.5 H_{p}$, which corresponds
well to a horizontal wavenumber of $\sim$ 15-30, approximately the
horizontal wavenumber at which we see a break in slope in
Figure~\ref{fig:comp-omkspec}.  This also explains why there is a time
variation in the break point between the slopes: at different times
plumes penetrate greater or lesser distances and are present to
greater or lesser extent hence, smaller and larger horizontal
wavenumbers where the spectral slopes change.  Since plumes likely
do not penetrate much further than the penetration depth they likely
dont contribute much to the smaller scales of Region B.  Therefore,
those waves must be predominantly driven by eddies at the interface.

We have seen in Figure~\ref{fig:plume-eddie-diagram} that eddies {\it
  and} plumes generate low frequency (low phase angle) waves.  Hence,
Region C has contributions from both plumes and eddies.  On the other
hand, in that same figure, and in animation, we see that plumes are
dominantly responsible for high frequency (high phase angle) waves.
The break in slope in the frequency spectrum occurs over a broad range
$\sim 10-80\mu Hz$.  If eddies are assumed to have length scales of
$l_{edd}$ and speeds of $v_{edd}$, then the typical eddy frequency is
$\omega_{edd}\sim v_{edd}/l_{edd}$.  We observe eddies generated at
the interface to have length scales of $\sim 0.2D_{cz}$ and speeds in
the range of $10^{5}$ cm s$^{-1}$,
the frequency associated with such eddies is $\approx 25\mu$Hz,
allowing for variations in speeds and eddie size, we gather that
turbulent eddies can generate frequencies up to approximately this
value, give or take a factor of a few.  We see the upper end of this
range is near the break point frequencies seen in
Figure~\ref{fig:comp-omkspec} and
Figure~\ref{fig:plume-eddie-diagram}.

On the other hand, if we assume plume frequencies are associated with
the typical time of a plume incursion, plume frequencies can be
estimated as $\omega_{pl}\sim v_{pl}/\Delta_{p}$, where $v_{pl}$ are plume
velocities, typically a few times larger than the eddie velocities.  With $\Delta_{p}$ the penetration depths
given in Section 3, we find frequencies in the range of 80-500$\mu$Hz.
The lower end of this range is similar to the position of the break
points in frequency.  Note that both of these frequencies are
proportional to some velocity at the convective-radiative interface,
and that these velocities decrease with increasing rotation rate.
This may explain why the break point in the frequency spectrum moves
to lower frequencies for higher rotation rates.

\begin{figure}
  \centering
  \includegraphics[width=4in]{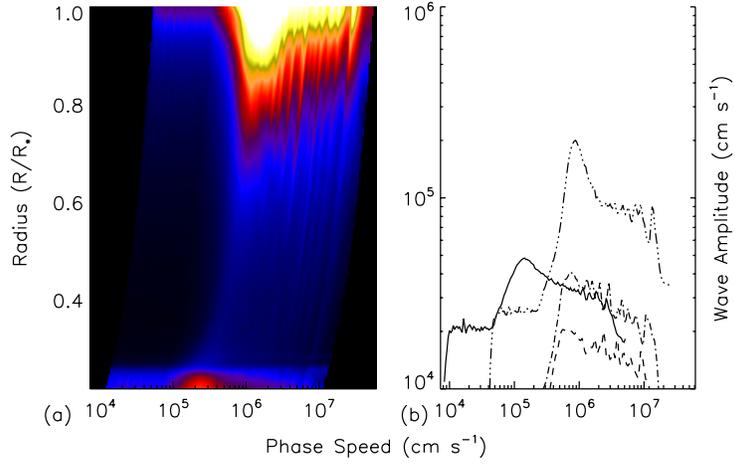}
  \caption{(a) Wave amplitude (in cm s$^{-1}$) as a function of phase
    speed and radius for model U5.  White represents high amplitude,
    while blue represents low amplitude.  (b) Horizontal slices
    through (a) solid line represents wave amplitude as a function of
    phase speed at a radius of 0.17$R_{*}$, dashed line 0.65$R_{*}$,
    dot-dashed line 0.79$R_{*}$ and triple-dot dashed line
    0.99$R_{*}$.}
  \label{fig:spec-phasespeed}
\end{figure}

Another way to investigate the wave generation process is to consider
the amplitudes of waves generated as a function of phase speed rather
than as a function of frequency and wavenumber individually.  This is
shown in Figure~\ref{fig:spec-phasespeed}.  There we see that IGW are
generated in a very narrow range of horizontal phase speeds from $1-4
\times 10^{5}$ cm $s^{-1}$, with a sharp cutoff at lower phase speeds
and a gradual decline at higher phase speeds.  In
Figure~\ref{fig:spec-phasespeed} we show the velocity amplitude, which
shows the same behavior as in Figure~\ref{fig:raddamp}, with
amplitude initially large, then decreasing throughout the bulk of the
radiative envelope and finally increasing again toward the surface.
This is distinctly different than the {\it energy} which would
decrease strictly in radius.  Moving outward into the radiative
envelope we see that the peak velocity amplitude occurs at higher and
higher phase speeds.  This is the result of both dissipation of the
lower phase speed waves, as well as energy transfer to higher
frequency waves (see Section 5.4).  While the energy is predominantly
in propagating IGW we also see peaks at distinct phase speeds in
Figure~\ref{fig:spec-phasespeed}b indicative of the standing g-mode
component.

It is noteworthy that the phase speed at which waves are predominantly
generated is nearly identical to the dominant radial velocity at the
interface (10$^{5}$-10$^{6}$ cm$s^{-1}$).  While the frequency and
wavenumber spectra of generated waves is broad, the phase speed
spectrum is relatively narrow.  Given the frequency and wavenumber
spectrum generated we could have seen a range of phase speeds between
$\approx 10^{4}-10^{6}$ cm s$^{-1}$, yet we find a much narrower range
over which energy is generated.

\begin{figure}
  \includegraphics[width=3.5in]{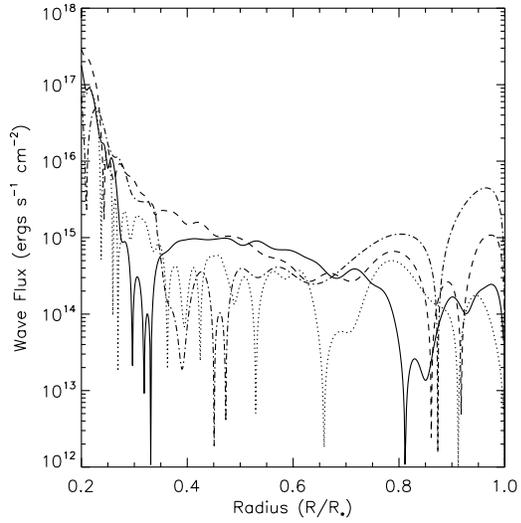}
  \caption{Wave energy flux as a function of radius for models U1
    (solid line), U2 (dotted line), U8 (dashed line) and U9
    (dash-dotted line).  While there are small variations in wave flux
    at generation, there are substantial variations at the surface,
    with higher rotation rates showing higher surface amplitude.}
  \label{fig:waveflux-pu}
\end{figure}

Another important quantity associated with wave generation is the
total flux delivered from the convecting region to waves.  This is
typically measured by the quantity:
\begin{equation}
  F_{w}=\overline{p'\vec{v}}
\end{equation}
which refers to the wave energy flux as derived in \cite{lighthill} where p' is the
pressure fluctuation and $\vec{v}$ is the velocity.  We show this quantity as a
function of radius for a number of our models in Figure~
\ref{fig:waveflux-pu} (see caption for details).  It has been
predicted in several studies \citep{gk90,ktz99,lecq12,shiodeq12} that
the integrated wave flux is
\begin{equation}
  F_{w}\sim M_{c} F_{conv}
\end{equation}
where $M_{c}$ is the convective Mach number and $F_{conv}$ is the
convective flux.  The typical radial velocities at the convective-radiative
interface are $\approx 10^{5}$ cm s$^{-1}$ with intermittent
excursions to much larger values.  The sound speed at the interface is
$\approx$ 5 $\times$ 10$^{7}$ cm s$^{-1}$, therefore, the convective
Mach number is $\approx 10^{-3}-10^{-2}$.  The total flux at the
convective-radiative interface is $\approx 10^{19}$ ergs
cm$^{-2}$s$^{-1}$, from Figure ~\ref{fig:waveflux-pu} we see that the
wave flux there is $\approx 10^{17}$ ergs cm$^{-2}$ s$^{-1}$, so the
ratio of wave flux to total flux is $\approx 10^{-2}$, but can be larger.  This is
generally consistent with the theoretical prediction encapsulated in
Equation (15).

In summary, we find that the spectrum of waves generated by
overshooting convection is broad, but generally with a decreasing
function of frequency and wavenumber.  Although we attempted many
functional fits to this data, we find that combined power laws fit the
data the best.  We see breaks in the power law fits in both frequency
and wavenumber, which could be attributed to the different physical
processes of wave generation by impeding plumes and convective eddies
at the interface.  These spectra are substantially different than
those predicted by the theories described in the Introduction.  When one considers
our ``low'' frequencies which generally span up to around 60-80$\mu
Hz$ the power law is $\sim -1$, {\it substantially} less steep than
the -4 - -6.5 power law predicted by \cite{ktz99} and \cite{lecq12}.  Similarly,
for our ``low'' wavenumbers which include wavenumbers out to 15-30,
the slope is $\sim -2$, also substantially different than that
predicted.  This is likely due to a number of effects, most
fundamentally that we include the effects of both impinging plumes and
eddie generation and that these generation processes interact with
each other, as do the waves they generate.  This interaction
  would likely lead to a flatter spectra, such as the one we produce.
While the total wave flux
is consistent with previous theoretical predictions our wave energy is
distributed more broadly in frequency and scale, showing less severe
drops in energy at higher frequency and wavenumber than previous
analytic predictions.  Possibly more fundamentally, rather than
  matching frequency and scale of convection individually we find that waves
are generated in a narrow range of phase speed, which corresponds to
the radial velocity of convective motions at the interface.
It is likely that these spectra will change somewhat in three
dimensions (3D).  The third dimension may provide another avenue for
transfer of energy to smaller scales and therefore, allow for a
steeper spectrum than the one we see here.  Probably more
fundamentally, it is likely that the nature of the turbulence and
plume dynamics changes.  It is difficult to say how these changes will
affect the wave spectra.

Until 3D numerical simulations of wave excitation can be done it might
be possible to learn something from experiments of IGW driven by
plumes and convection.  Several experiments of turbulence and plume
driven waves have been conducted \citep{ansuth10,holdsworth12} and
have found that these processes generate waves in a narrow range of
frequency, $0.4\leq \omega/N \leq 0.7$.  The physical reason for this narrow
range is still unknown, although the authors note that these values of
the frequency are optimal for transporting horizontal momentum
vertically and speculate that there is a nonlinear interaction between
the driving region and wave propagation region in order to bring about
maximum angular momentum transport.  If this is true, it would imply
that the simple theories of wave excitation discussed in the
introduction are inadequate as they completely neglect any wave
influence on the convection.  More recent experiments of convectively driven IGW
  are being carried out by M. LeBars at the SpinLab at UCLA,
  preliminary results show a power law in frequency similar to those
  found in our simulations (private communication), however more
  quantitative comparisons await more complete data from the
  experiments.  How these experimental results extend to stars
with a rapidly varying Brunt-Vaisala frequency is unknown.  

\subsection{Wave Propagation}
Once generated at the convective-radiative interface, IGW propagate
toward the stellar surface.  As discussed in Section 5.1 the wave
propagation is outward, along spiral paths, with the angle of phase
lines with respect to the horizontal set by the ratio of $\omega/N$.
As waves propagate toward the surface their amplitudes are affected by
two main effects, radiative dissipation and conservation of
pseudo-momentum.  By conservation of pseudo momentum $(\overline{\rho
  v_{\phi}'v_{r}'})$, the wave amplitude increases like
${\rho}^{-1/2}$.  In these models the density decreases from
approximately 30 gm cm$^{-3}$ at the top of the convection zone to
$10^{-3}$ gm cm$^{-3}$ at the top of the radiation zone, so one would
expect the wave amplitudes to increase by nearly two orders of
magnitude.  This acts on all waves, regardless of frequency and
wavenumber.

Second, waves are radiatively dissipated.  Because gravity waves are
thermal perturbations they are subject to radiative dissipation.  The
wave amplitude, acted on by the combined action of density
  stratification and radiative dissipation can be written as:
\begin{equation}
  A(\omega,k_{h},r)=A(\omega,k_{h},r_{0})\overline{\rho}^{-1/2}e^{-\tau(\omega,k_{h},r)}
\end{equation}
where
\begin{equation}
  \tau(\omega,k_{h},r)=\int\limits_{r_{0}}^{r} \kappa k_{h}^{3}\frac{N^2}{\omega^{3}} dr
\end{equation}
where $r_{0}$ is the radius at the interface and
$A(\omega,k_{h},r_{0})$ is the wave amplitude at the interface.  By
Equation (17) radiative dissipation is most effective at dissipating
low frequency, high wavenumber (small scale) waves.  Which can be
physically understood as short wavelengths have high curvature and are
therefore, dissipated rapidly.  Similarly, short frequencies have a
longer time over which to feel the effects of dissipation.
Alternatively, using the relationship $\omega/N=k_{h}/k_{v}$, the
integrand in Equation (17) can be rewritten as $\kappa k_{v}^{3}/N$
and one can see that waves with short vertical wavelength are dissipated more
efficiently.  

\begin{figure}
  \includegraphics[width=4in]{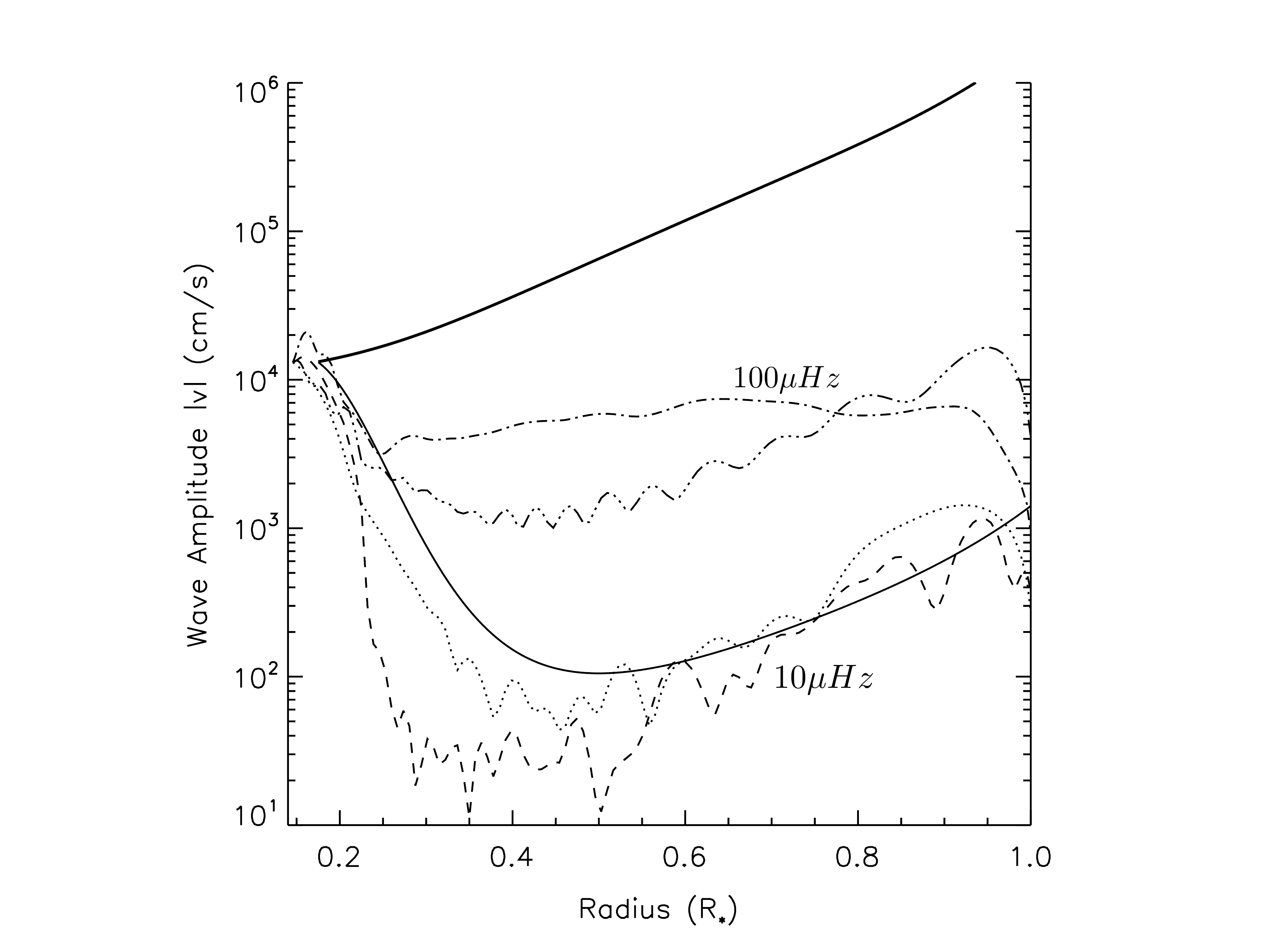}
  \caption{Wave amplitude as a function of radius calculated
    analytically using Equations (16) and (17) (solid lines) and
    calculated from the numerical simulations of D11 (broken lines
    with clear wavelike patterns).  Three solid lines at top overly
    each other and represent theoretical predictions for $k_{h}$=1
    with frequency 10$\mu$ Hz and 100$\mu$ Hz waves respectively, and
    for $k_{h}$=10 with frequency 100$\mu$Hz.  Solid line at bottom
    represents theoretical prediction for $k_{h}$=10 and
    $\omega$=10$\mu$Hz.  Wave amplitudes from the numerical simulation
    are represented as broken lines for $k_{h}$=1,$\omega$=10$\mu$Hz
    (dotted line), $k_{h}$=10, $\omega$=10$\mu$Hz (dashed line),
    $k_{h}$=1,$\omega$=100$\mu$Hz (dot-dashed line) and
    $k_{h}$=10,$\omega$=100$\mu$Hz (triple-dot dashed line).}
  \label{fig:raddamp}
\end{figure}

These two effects on the propagation of IGW are the most fundamental.
Clearly, in massive stars where the waves propagate into regions of
decreasing density, density stratification and radiative dissipation
have opposite effects.  Figure~\ref{fig:raddamp} shows these
combined effects for a couple of frequency and wavenumber
combinations, which was found by integrating Equations (16) and (17)
above using the thermal diffusivity set in the simulation and the
Brunt-Vaisala frequency from the stellar model, yet using amplitudes
at the interface given by the simulation (solid lines).  Also, shown
in that figure are the amplitudes of waves, as a function of radius,
from our numerical results for the same frequency/wavenumber
combinations (broken lines, easily discerned as waves, whereas the
model shows straight lines).  There we see that low frequency waves 
are dissipated similar to predictions, while high frequency waves
are dissipated more than theoretical predictions, with some dependence
on scale.  The discrepancy between the theoretical prediction and
numerical simulations is more severe for the lowest frequencies
and smallest scales.  Theoretically those waves would have virtually
no amplitude at the surface, while the numerical simulations show
amplitudes similar to those shown in Figure~\ref{fig:raddamp} for
higher frequencies and lower wavenumbers.  Overall, the numerical
simulations show far less dispersion between wave amplitudes of
different frequencies and wave numbers than the simple theory would
predict.  The difference between the predicted behavior and that seen
in the simulations is likely due to nonlinear effects such as
wave-wave interaction, wave-breaking or wave-mean flow interaction
(see Section 5.4).

\begin{figure}
  \includegraphics[width=4in]{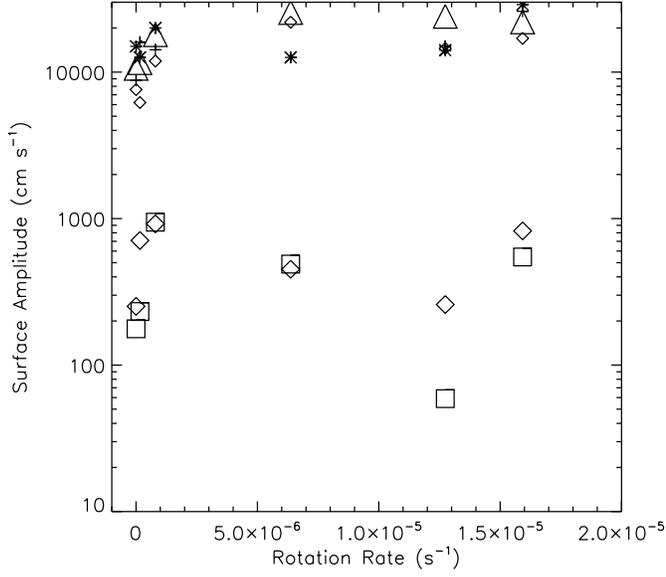}
  \caption{Surface amplitudes of selected waves as a function of
    rotation rate.  Small pluses, diamonds and asterisks represent
    maximum amplitudes at various times.  Large triangles represent
    the average of maximum amplitudes.  Note that all of these maximum
    values have high frequencies ($> 50\mu$Hz) and small scale ($k_{h}
    >8$).  Large diamonds and squares represent amplitudes for waves
    with a frequency of 10$\mu$Hz and $k_{h}=1,2$ respectively.}
  \label{fig:surfamp-rot}
\end{figure}

Figure~\ref{fig:surfamp-rot} shows the surface velocities for a subset
of our models and for a variety of frequency/wavenumber combinations
(see figure caption for details).  We often see waves which are nearly
(but not quite) a km/s.  The large triangles represent the maximum
velocities obtained during the times we investigated.  These maxima
are ${\it always}$ associated with higher frequency, smaller scale
waves.  This is not quite in the range of the non-thermal surface
velocities needed to explain macroturbulence \citep{aerts09}, but
considering the enhanced diffusivity and viscosity in these
simulations, it is possible to imagine IGW potentially contributing to
the ``macroturbulence'' necessary to explain spectral line widths
\citep{howarth97}.

Recently, \citep{shiodeq12} have investigated the detectability of IGW
in massive stars by the spacecraft mission Kepler.  In that study the
emphasis is on g-modes, the waves which form a standing pattern
between the surface and convective-radiative interface.  The waves
most likely observable would be those of the largest scale.  We show
large scale waves ($k_{h}=1,2$) of frequencies 10$\mu$Hz, as diamonds
and squares, respectively, in Figure~\ref{fig:surfamp-rot}.  There we
see that these waves have much lower amplitudes than the maximum
values, ranging from $50-10^{3}$ cm/s.  These amplitudes are
approximately an order of magnitude larger than those predicted by
\cite{shiodeq12}.  There are several likely reasons for this
discrepancy.  First, in our simulations high frequency waves are
excited directly by plumes, with fairly high amplitude, rather than
relying on the incoherent addition of multiple eddie sources.  Second,
\cite{shiodeq12} only consider standing waves while our waves likely
have a propagating and standing component (as seen in
Figure~\ref{fig:powspec_3heights_nv8}).  Finally, as discussed in
Section 2, we have a larger luminosity than an actual star of this
size, this results in a larger velocity within the convection zone,
although we believe this is offset by enhanced diffusion (see Section
6).

\subsection{Wave-Mean Flow Interaction}

As shown by \cite{ep60} IGW can only transport angular momentum if
they are attenuated.  Typical attenuation mechanisms include radiative
diffusion, wave breaking at critical layers or large amplitude wave
breaking.  In practice it is difficult to ascertain which dissipation
mechanism is at work under various circumstances.

\begin{figure}
  \includegraphics[width=4in]{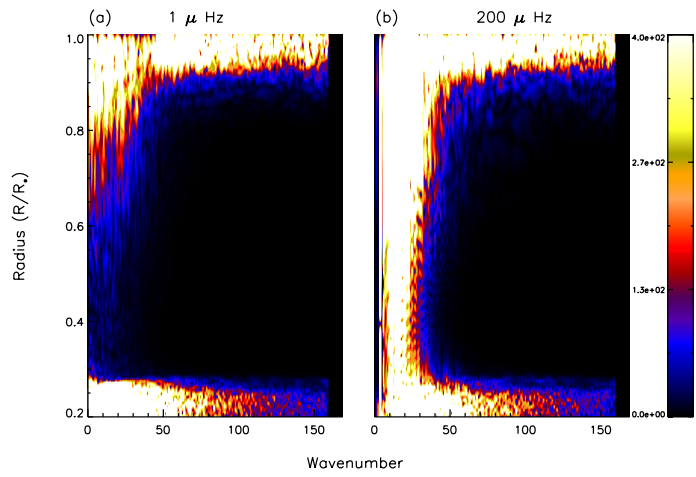}
  \caption{Wave amplitude at given frequencies as a function of wavenumber and
    radius, corresponding to a time of $\sim 2\times 10^{6}s$ in model
    D11.}
  \label{fig:enwavrad}
\end{figure}

We discussed the effects of radiative dissipation in the previous
section and here we turn to critical layer absorption and wave
breaking.  Figure~\ref{fig:enwavrad} shows the wave amplitude as a
function of wavenumber and radius for both high and low frequencies.
What we see is that wave generation is dominated by low wavenumbers
($\le 30$, as also shown above in
Figure~\ref{fig:powspec_3heights_nv8} and in Section 5.2).  We see
that only the largest scales and highest frequencies have appreciable amplitude within the
bulk of the radiative envelope but that near the surface high
amplitudes are seen even for small scales and low frequencies.
Furthermore, for the smallest scales the amplitudes at the surface are
larger than they were at generation.  This implies generation of small
scale waves ${\bf at}$ ${\bf the}$ ${\bf surface}$ at both low and
high frequencies.  Transfer of energy from large scales to small
scales occurs both during wave breaking as well as when waves
encounter critical layers \citep{rmac08}.

\begin{figure}
  \includegraphics[width=4in]{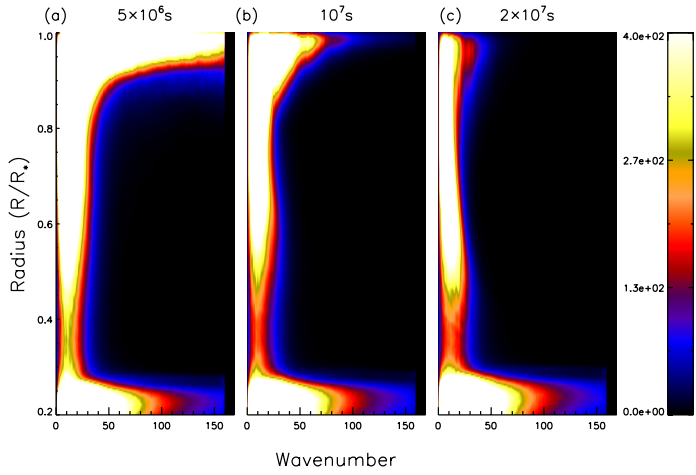}
  \caption{Energy integrated in frequency at various times, as a
    function of wavenumber and radius for D11.}
  \label{fig:enwavrad-int-time}
\end{figure}

If we look at the same plot, but at various times, integrated in
frequency, as in Figure~\ref{fig:enwavrad-int-time}, we can
distinguish between critical layers and typical high amplitude wave
breaking.  In Figure~\ref{fig:enwavrad-int-time}a we see that there is
substantial energy at small scales at the surface.  This time is
sufficiently early that a strong mean flow has not yet developed and
there is, as yet, no critical layer
(Figure~\ref{fig:angvel-nv8-nv14}).  Therefore the energy at
smaller scales is most likely due to high amplitude wave breaking.
Figure~\ref{fig:enwavrad-int-time}b represents a time around when the
critical layer is forming (also see
Figure~\ref{fig:angvel-nv8-nv14}).  At this time energy is still
being transferred to small scales by wave breaking but some of that
small scale energy is being absorbed into the critical layer (smaller
scale waves have smaller phase speeds and are therefore, more
susceptible to critical layer absorption).
Figure~\ref{fig:enwavrad-int-time}c shows the wave energy after the
mean flow has developed, we see that there is only energy in the
largest scales (probably predominantly in standing g-modes).  Energy
in smaller scale waves, that was generated by wave breaking is now
being absorbed by the critical layer.  Note that once the critical layer develops
the linear velocity at the surface is on the order of $\sim
10^{6}-10^{7}$ cm s$^{-1}$, precisely equal to the phase speed where
IGW are peaked at the surface (Figure~\ref{fig:spec-phasespeed}).

The wave-mean flow interaction we have described above is very
complex.  It involves both wave breaking and critical layer
absorption, the development of a mean flow which then alters the
properties and propagation of the waves and the subsequent evolution
of the mean flow.  This all occurs from the propagation and
interaction of a broad spectrum of convectively generated waves in a
highly stratified gas.

\subsection{Angular Momentum Transport}

\begin{figure}
  \centering
  \includegraphics[width=4in]{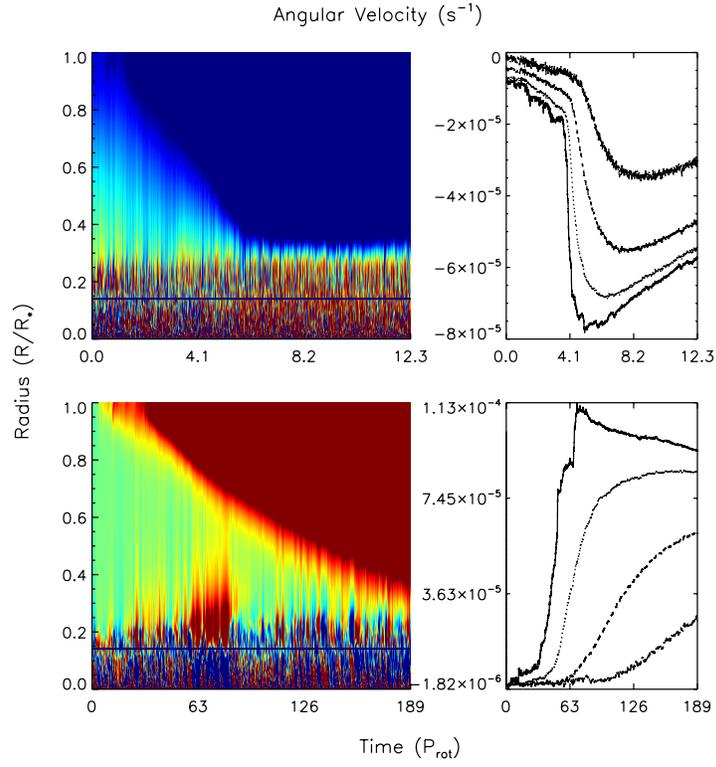}
  \caption{Left panels: Angular velocity as a function of time and
    radius for models U2 (top) and U8 (bottom).  Right
    panels show horizontal cuts through figures on the left, thus
    showing the angular velocity as a function of time for several
    radii, corresponding to 0.96R$_{*}$ (solid line), 0.82R$_{*}$
    (dotted line), 0.61R$_{*}$ (dashed line) and 0.32R$_{*}$
    (dot-dashed line).}
  \label{fig:angvel-nv8-nv14}
\end{figure}

In Figure~\ref{fig:angvel-nv8-nv14} we show the angular velocity
over time for three of our models (U2 (top) and U8
(bottom)).  There we see, as described above, a mean flow develops
first at the stellar surface.  Over time the angular velocity migrates
toward the source of the waves (the convection zone) in time.  As it
does so the convection zone starts to spin predominantly with the
opposite sense, in order to accomodate angular momentum conservation.
We see that the most rapidly rotating model, U8, achieves a higher
absolute angular velocity at the surface.  In general, we find that
the extreme angular velocities seen in
Figure~\ref{fig:angvel-nv8-nv14} bracket the maximum values
achieved in our models and that the trend is for faster rotators to
achieve slightly higher mean flow amplitudes.  This is not a strong
effect as a factor of 100 increase in rotation leads to an $\approx$
40\% increase in angular velocity.  

 One will notice that the modulation of angular velocity in time
  is not symmetric.  The growth is initially dominated by critical
  layer absorption, which causes a {\it rapid} increase in angular
  velocity.  Once the angular velocity is saturated, the amplitude is
  so large that it is not critical to many waves, so the decay is much
  slower, being dictated by a weaker effective wave flux and viscous
  dissipation of the mean flow.
  
\begin{figure}
  \includegraphics[width=4in]{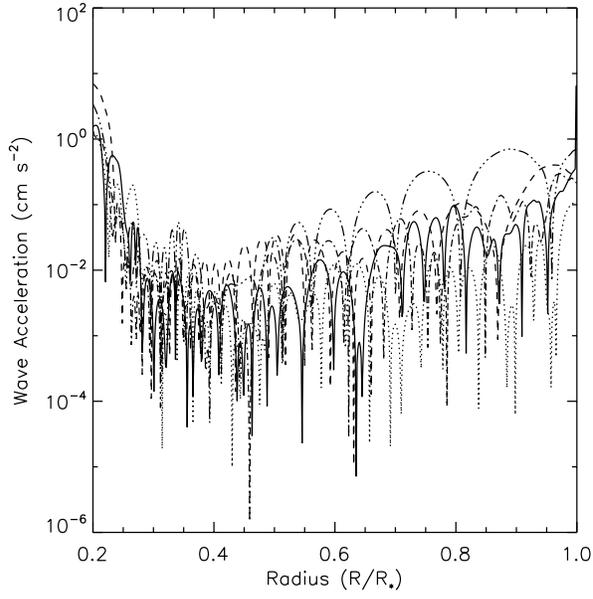}
  \caption{Second term on left-hand side of Equation (8) averaged in
    time as a function of radius for models U1 (solid), U2 (dotted),
    U5 (dashed), U8 (dot-dashed) and U9 (triple dot-dashed).}
  \label{fig:wave-acceleration-av}
\end{figure}

The mean angular momentum transport by IGW is described by Equation
(8), $\overline{v_{r}v_{\phi}}$ is often referred to as the
wave flux and the second term on the left hand side of Equation (8) is
the acceleration of the mean flow due to waves, we refer to it as wave
acceleration.\footnote{Viscous diffusion terms are negligible {\it
    most} of the time and at {\it most} locations.}  In Figure~\ref{fig:wave-acceleration-av} we show the time average of
the wave acceleration as a function of radius for several of our
models.  There we see that wave acceleration is substantial both near
the base and at the top of the radiative envelope.  With accelerations
nearing 0.1-1 cm s$^{-2}$ the linear (and angular) velocity at the
surface can change dramatically on relatively short timescales.  For
example, with this acceleration, an initially non-rotating star can
achieve surface angular velocities of 10$^{-5}$ -10$^{-4}$ rad
s$^{-1}$ in a few years.  This is consistent with the accelerations
seen in Figure~\ref{fig:angvel-nv8-nv14}.

Upon inspection of Figure~\ref{fig:angvel-nv8-nv14} one will see
  that angular velocities increase by a factor 10 in a few rotation
  periods (depending on the model).  The angular velocities attained are a function of the wave
  flux, which itself is a function of the convective flux, set by
  nuclear energy generation.  In our simulations the convective flux
  is larger than the stellar convective flux by a factor $\sim 5\times
  10^{4}$ to offset our larger than stellar thermal diffusivity.  By
  mixing length theory this would
  lead to velocities which are $\sim$30 larger than stellar convective
  velocities.  If these velocities were {\it not} offset by the larger than stellar
  diffusivities, then the wave flux, as
  described above would be $\sim$ 1000 too large.  Dunkerton (1981)
  has argued that the period of oscillation is inversely proportional to the
  wave flux.\footnote{\cite{plumb77} argued that
    the period of oscillation was inversely proportional to the
    product of the wave flux and damping rate.  However, this was for Boussinesq
    systems.  In compressible systems \cite{dunk81} argued, and his
    compressible simulations confirmed, that there was very little
    dependence on the
    damping rate.  \cite{rmac08} also saw little variation in
    oscillation period with the thermal diffusivity.}  Using this scaling, our periods of modulation are too
  short by a factor of $\sim$1000.  Therefore, if we apply
  this scaling we arrive at a modulation time of $\approx$
  a few-10000 rotation periods.  For our fast rotators this results in
  modulation times of 10-30 years.  This estimate comes with some
  caution.  First, it is difficult to estimate how much the wave
  fluxes are enhanced.  We believe that our large wave driving is
  offset by our large wave damping, so a factor of 1000 is likely an
  overestimate.  Second, the scaling of the period
  of oscillation has come from simplified models with few waves and no
  critical layers.  Finally, the timescale for modulation will depend
  on the phase in which one is observing.  During the critical layer
  growth phase, variations occur much more rapidly, so one may see
  variation in a few years.  However, during the decay
  phase, when no critical layers are present and net wave fluxes are
  low, modulation may occur on a longer timescale.

\section{Dependencies}
Wave generation, propagation, dissipation and consequent angular
momentum transport could depend on the parameters of our simulations.
In this section we discuss how various parameters affect the most
important aspects of our results: spectra of wave generation including
amplitudes and consequent angular momentum transport.  We will also
discuss other dependencies, such as convective velocities and
convective overshoot as they pertain to the wave spectra and angular
momentum transport.

\subsection{Rotation}

We see in Figure~\ref{fig:comp-omkspec} that the wave spectra varies
some with rotation rate.  In general, these variations are not larger
than variations within a given model in time.  However, there are a
couple of clear trends.  First, the low frequency wave spectra are
flatter as the rotation rate is increased.  This occurs because higher
rotation rates allow for larger differential rotation.  Larger
differential rotation leads to larger doppler shifts of wave
frequencies, and therefore, a larger range of wave frequencies,
particularly at low frequencies.  This gives rise to a flatter
spectrum at lower frequencies.  This is also true at higher
frequencies, but higher frequencies are shifted less, relative to
their fiducial frequency, so the effect is weaker, as is seen in the
spectra.  Second, the frequency at which the slopes of the spectra
change tends to be lower for higher rotation rates.  This is likely
because radial velocities are decreased in rapidly rotating models.
Since radial velocity is directly related to the plume intrusion
depth, this affects the frequencies of waves generated by plumes and
hence, the break point frequency.  This also explains the shift in
wavenumber spectra.

At our highest rotation rates, the rotation frequency is $\approx
16\mu Hz$.  This is very similar to the frequency at which we see a
pile up of wave energy.  This behaviour has been seen in experiments
as well \citep{holdsworth12}.  The Rossby number which is a measure of
the inertial terms to the rotation terms is defined as
\begin{equation}
  R_{o}=\frac{W_{o}}{fR}
\end{equation}
where R is the radius at the interface, $W_{o}$ is a typical radial velocity
at the interface and the other variables as
defined above, we see that our fastest rotation rate corresponds to a
Rossby number of $\sim 0.85$.  This is the only model of ours for
which the Rossby number is lower than 1, indicating the Coriolis
forces dominate the inertial forces.  This may be why this is the only
model to show a distinct buildup at the inertial frequency.  It may
also be because this is the only model for which the inertial
frequency is at a high enough frequency to not be overwhelmed by low
frequency waves driven by plumes and eddies.

We also find that, on average, wave amplitudes, mean flows and wave
acceleration are all {\it slightly} larger in models with rapid
rotation than in more slowly rotating models.  The energy equation is
the same in a rotating model, as the Coriolis force does no work.
However, in rotating systems, energy is not partitioned equally
between kinetic energy and potential energy (due to stratification).
In rotating systems, more of the total energy is in the form of
kinetic energy.  In fact, the ratio of kinetic energy to potential
energy is directly related to the rotation rate \citep{gill82}
\begin{equation}
  \frac{KE}{PE}=1+\frac{\Omega^{2}}{N^{2}}tan^{2}\phi
\end{equation}
this means that in rotating systems wave velocities can be larger than
their non-rotating counterparts at the expense of potential energy in
the stratification.  This effect is small, but it is noticeable and
explains the small variations seen in wave amplitudes, wave
acceleration and resulting mean flow.

\subsection{Differential Rotation}

There is little variation with any parameters when differential
rotation is considered.  We find a slight variation in penetration
depth, as to be expected, as differential rotation produces a
Kelvin-Helmholtz instability causing additional mixing.  For
differentially rotating models the spectra are more consistent
and slightly flatter in frequency and smaller scales.  This is consistent with
these models being slightly more turbulent in the overshoot region due
to shear instability.  However, overall the variation between these
models and the initially uniformly rotating models is not substantial
compared to the overall variation of spectra in time and with rotation
rate.  Models with large differential rotation develop strong mean
flows, while the one model with weak initial differential rotation
does not, so it is likely that initially differentially rotating
models develop strong mean flows more readily and that those flows
have a preferential sense. Overall, the differentially rotating models
are similar to those with no initial differential rotation.

\subsection{Diffusivities and Heat Flux}

Besides rotation, the other tunable parameters are the thermal
diffusivty, $\kappa$ the viscous diffusivity, $\nu$ and the heat flux,
$\overline{Q}$ (or convective driving).  Taken together these numbers,
in the appropriate non-dimensional form, constitute the Rayleigh
number:
\begin{equation}
  Ra=\frac{g\overline{Q}D^{5}}{c_{v}\kappa^{2}\nu\overline{T}}
\end{equation}
which dictates the details of the convection, but does not necessarily
dictate angular momentum transport by IGW.  Regardless, we would like
this number to be large to adequately represent turbulent motion
within the convection zone.  From the values listed in Table 1, we see
that all of our models have extremely high Rayleigh numbers ($\sim
10^{14}$ by this definition), except possibly model MU1-4.

Another control parameter that affects convection (and possibly waves)
is the Prandtl number (Pr=$\nu/\kappa$).  In stellar interiors this
number is substantially smaller than 1 ($\sim 10^{-7}$).
Unfortunately, these values are unobtainable in numerical simulations,
especially while maintaining large Rayleigh numbers.  Nearly all of
our simulations have Pr larger than 1, except model MU1-4.  In that
model we achieve a Pr less than 1 by increasing $\kappa$, but this
leads to extremely laminar flows (the Rayleigh number is reduced)
which we think are less realistic than our turbulent flows with
inaccurate Prandtl numbers.  We argue that as long as our viscous
dissipation is low enough to not dominate the momentum equation proper
dynamics are recovered.

While the above parameters dictate the details of convection, what we
are primarily concerned with here is what parameters affect the wave
dynamics.  Like convection, we expect all of these parameters to
affect waves and wave driven momentum transport.  Precisely how these
parameters affect this transport is what we address here.  First,
convective forcing, in the form of $\overline{Q}$, determines the
velocities within the convection zone and therefore, it affects the
vigour of convection (the Rayleigh number).  Second, the thermal diffusivity dictates the diffusion of waves in
their propagation to the surface and therefore, affects wave
amplitudes and momentum transport.  Finally, the viscous diffusivity
affects the mean flow evolution.

We first address the spectra of waves generated.  The mean spectral
slopes of the modified models are (neglecting model MU1-3) $m\sim
0.72\pm 0.18$ for low frequencies and $m\sim 4.16\pm 0.17$ for high
frequencies and $n\sim 1.66\pm .05$ at large scales and $n\sim 4.86\pm
0.55$ at small scales.  These slopes are very similar to that of model
U1 and certainly no more exceptional than the time variation within
that model.  Therefore, we conclude that none of these parameters
affect the spectra much, with the exception of model MU1-3.  In that
model, the thermal diffusivity is so high that it dominates buoyancy
and therefore, dissipates waves extremely efficiently, leading to
steep slopes in the wavenumber spectra.  Model MU1-2 which has a
slightly lower thermal diffusivity shows virtually no difference in
spectral slope (compared to the fiducial model U1).  While it is
possible that substantially lower thermal diffusivities could lead to
flatter spectra, we expect (hope) that we have achieved large enough
Rayleigh numbers that the buoyancy forces dominate diffusion enough so
that the spectra remains relatively unchanged, this is somewhat
validated with MU1-2.  However, when we increase $\kappa$ enough
(MU1-3) this no longer becomes valid and the spectra changes
drastically.

It is difficult to measure how much Q, $\kappa$ and $\nu$ affect the
surface amplitudes of the waves beyond basic qualitative
understanding. Comparing the amplitudes of the largest scale waves
($k_{h}=1$) and frequencies of 10$\mu$Hz of models U1 with MU1-2
(which we expect should have similar velocities if our scaling
described in Section 2 is accurate), we find that on {\it average} the
velocities are similar (average radial velocity for U1 is ~290 cm
s$^{-1}$, while for MU1-2 it is 240 cm s$^{-1}$).  However, we note
that both models vary substantially (by as much as an order of
magnitude) over the extent of the simulation.  This isnt too
surprising given the variation in the mean flow and its ability to
absorb wave energy.  The time variation of the mean flow, and its
affect on wave amplitudes makes direct comparisons difficult.
However, we note that, on average, MU1-1, MU1-3 and MU1-4 all have
lower velocity amplitudes at the surface compared to U1, while MU1-6
which has a lower Q and a similarly lower $\nu$ has higher velocity
amplitudes (by a nontrivial margin).

When there is sufficient driving to offset dissipation, the slopes of
the spectra as shown in Table 1 depend primarily on the time and
spatial correlations of waves.  However momentum transport by waves
depends on a combination of the wave driving amplitude (the
convection, $\overline{Q}$) and how those waves are dissipated during
their propagation.  Most fundamentally, it must be recognized that in
order for a mean flow to develop the wave acceleration must be larger
than the viscous diffusion (Equation 8), so reductions in
$\overline{Q}$ and increases in $\kappa$ if sufficiently large could
lead to wave accelerations which are too small compared to the viscous
dissipation to allow a mean flow to develop.  This is just a statement
that, in order for a mean flow to develop, the local Reynolds number
$=v_{r}L/\nu$ must be greater than 1.  This is most certainly the case
in stellar interiors.  Even with conservative estimates of wave speeds
and scales \citep{shiodeq12} the Reynolds number at the stellar
surface should be substantially larger than 1 (assuming viscosities of
order $\sim 10$ gives Reynolds numbers of $\sim 10^{7}$) .  However,
in numerical simulations this parameter space is difficult to achieve.

Given the velocities and scales seen in Figure~\ref{fig:surfamp-rot}
(squares and diamonds) combined with the viscous diffusivity listed in
Table 1, the Reynolds number in our standard simulations (``U'' and
``D'' models) is $\sim$ 1-10.  Therefore, any reduction in
$\overline{Q}$, increase in $\kappa$ or $\nu$ is likely to shut off
mean flow acceleration.  Indeed we find this is the case, as models
MU1-1-MU1-5 all show decreased Reynolds numbers and (substantial) mean
flows do not appear within the (rather long) time we ran the
simulations.  However, model MU1-6 which has reduced driving
($\overline{Q}$), but similarly reduced $\nu$ (hence, the Reynolds
number should still be $>$1) does show mean flow acceleration.

\section{Discussion}

These results have many theoretical and observational consequences, a
few of which we discuss here.

1) {\it Misalignment of Hot Jupiters around Hot Stars} observations
have important implications on the origin of these hot Jupiters,
especially since a modest fraction of them show signs of significant
obliquity angle $\Theta$.  At face value, this observation would
disfavor the disk-migration scenario compared to a dynamical mechanism
for the origin of hot Jupiters.  However, hot Jupiters with high
$\Theta$ are primarily found around stars with effective temperature
($T_\ast$) hotter than 6,250K \citep{winn10}, while the orbits of hot
Jupiters around relatively cool, solar-like, stars are generally
aligned with their hosts \citep{winn10,schl10}.

Several dynamical mechanisms including Kozai resonances, planetary
encounters, and secular chaos have been suggested as possible
causes for the excitation of extra solar planets' obliquity and
eccentricity \citep{fab07}.  However, these processes do not
explicitly depend on the mass of the star, $M_\ast$.  Therefore,
\citet{albrecht12b} have suggested the following scenario leading to
misalignment: 1) hot Jupiters around all stars attained a random
obliquity distribution when they relocated to stellar proximity
under one of the dynamical mechanisms discussed above, 2) their
obliquities then evolved through tidal dissipation, on a time scale
$\tau_\Theta$, and 3) the tidal dissipation timescale $\tau_\Theta$ is
shorter in cool stars because of their convective envelopes, thus
$\Theta$ is correlated with stellar type because planets around cool
stars have had time to align through tidal dissipation, while planets
around hot stars have not.

The above interpretation is based on the application of a model of the
stellar tide by \citet{zahn77}, the quantitative details of which
remain uncertain \citep{ogilvie07, goodman09, barker10}.  If one
resorts to the equilibrium tide for dissipation of obliquity, as in
\cite{alb12}, the semi-major axis is simultaneously damped and the planets will fall
into their host stars.  On the other hand, if one relies on the
dynamical tide introduced by \citet{lai12} then planets evolve to
aligned or anti-aligned systems, contrary to observations.  Therefore,
there currently appears to be no tidal mechanism which can
explain the observations \citep{rl13}.

Several other scenarios have been suggested as potential causes for
the observed obliquity misalignment that are still consistent with the
disk migration scenario.  They can be categorized in terms of
reorientation of either the protoplanets' natal disks
\citep{bate10,lai11,baty12} or the stellar spin \citep{rll12}.  For
example, the spin orientation of protostellar disks may re-orient
episodically, especially during the disk formation \citep{bate10}. If the host stars'
spin and the planets' natal disks' angular momentum vector are not
initially parallel, an induced magnetic field (due to the interaction
between stellar magnetosphere and the disk) could introduce a torque
which will have the tendency to amplify the misalignment
\citep{lai11}.  In general, the ``tilted-disk'' scenarios do not
provide obvious explanations for the observed correlation between
obliquity and stellar mass.  In contrast, the ``stellar spin
modulation'' scenario described here and in \cite{rll12} relies on the angular momentum
redistribution due to the excitation, propagation, and dissipation of
IGW.  These processes are only effective in intermediate and massive
stars with convective cores and radiative envelopes.

Based on the results presented in the previous sections, we anticipate
the following observable tests of ``stellar spin modulation''.  A) The
surface rotational speed and orientation may modulate on a time scale
$\sim 10^4$ internal spin periods (or less), a timescale which should
be observable.  B) Since the surface rotation velocity change is due
to the internal angular momentum redistribution, we anticipate radial
differential rotation in these hot host stars, which may eventually be
measureable with asteroseismology \citep{jcd11}.  We furthermore
expect there to be latitudinal differential rotation which would
affect modeling efforts of the Rossiter-McLaughlin effect.  C) We
anticipate the spin of some intermediate-mass host stars may be
misaligned with their multiple transiting planets (which are
essentially coplanar).  In contrast, the spin of solar type host stars
may be well aligned with the orbits of multiple planetary systems.  D)
If detected, we expect that even long-period isolated planets around
intermediate or massive stars could show spin-orbit misalignment,
while those around solar type stars would show spint-orbit alignment.
E) Isolated massive stars may show modulation of their observed
rotation, v $\sin${\it i}.

The verification of A) or D) would invalidate, and C) would challenge
the "dynamical" models for the origin of planetary obliquity.
Observation of an obliquity variation in time (A) would invalidate and
(D) would challenge the ``tilted disk'' models.  On the other hand,
our IGW model is not able to account for the misalignment between the
orbits of hot Jupiters with the spins of their solar type host stars.
Some of those may indeed be due to dynamical effects, such as the
Kozai mechanism \citep{wm03}.

2) {\it Internal gravity waves in interacting binary stars.}  The
orbits of many short-period close binary systems, with
intermediate-mass or massive stars are observed to be circular despite
their short life span \citep{prim77}.  Their eccentricity $e$ appears
to decrease with the ratio $r$ of the stellar radius to the binary
separation \citep{giuricin84} and the inverse of their orbital period
$P^{-1}$.  In addition, their degree of synchronization (measured from
the ratio of stellar spin periods, $P_\omega$, to the binary systems'
orbital periods $P_b$) also rapidly decreases with $r$ toward unity
for $r >0.05$ \citep{giuricin84}.

These correlations provide supporting evidence to the dynamical
models \citep{zahn75,zahn77,sp83} based on the assumption that angular
momentum is effectively transferred between the stellar spin and
binary's orbit through the excitation of gravity waves at the boundary
of the convective-radiative interface by tidal perturbations,
propagated through the radiative envelopes and dissipated through
radiative damping at the surface.  These waves have the relative
angular frequency between the stellar spin and binary orbit.  Most
massive stars are born with spins faster than the orbital frequencies
of binary stars so that the tidally induced gravity waves generally
carry a negative angular momentum flux. \cite{gn89a,gn89b} postulated
that an angular momentum deficit is initially deposited in the
outermost regions of the radiative envelope and subsequently migrate
toward the convection zone.  After these regions have been despun to a
state of synchronism, tidally driven gravity waves are unable to
transport angular momentum\footnote{At this point the driven frequency
  is such that the group velocity of the waves tends to zero} and are
therefore, unable to force a sub-synchronous state.

The physical processes we have described in this paper are basically
analogous to these dynamical-tide models with the exception of the
excitation mechanism.  Internal gravity waves generated by convective
eddies or plumes have a broad range of frequencies and are therefore,
able to propagate and transport angular momentum in a range of
background rotation rates.  Depending on the rotation rate of the
convective cores, these waves may be able to propagate into the
surface regions which has already been synchronized by the binaries
tidal perturbation.  The dissipation of these waves may lead to
episodic modulation of surface spin rate around the state of
synchronization, including the possibility of sub-synchronous
rotation.  However, in binaries with relatively small $r$'s, the
degree of non synchronism is likely to be limited by the relative
strength of tidal forcing compared with convective forcing (which
still needs to be determined).

One interesting system is the interacting eclipsing binary system
which contains one of the best studied x-ray pulsars Her X-1 and a
$\sim 2 M_\odot$ main sequence star HZ Her which is overflowing its
Roche radius.  For the observed value of $r$ $(\sim 0.7)$, it is
natural to expect the tidally excited gravity waves to induce HZ Her
to spin synchronously with the binary's orbit.  Based on their
spectroscopic data, \cite{re97} placed an upper limit of 20\% on the
departure from a state of synchronism.  The results of combined x-ray
and optical data by \cite{ls98} also indicate sub-synchronous
rotation.  We suggest that internal gravity waves excited by HZ Her's
convective core may provide a possible hypothesis for this state.

Her X-1 is one of the first x-ray pulsing neutron stars to be
discovered \citep{gia73}.  Observations show three distinct periods,
one associated with the spin of the neutron star, one associated with
the binary orbital period and a third 35 day period \citep{ph87}.
\cite{gb76} have suggested the origin of this longer period is due to
the neutron star accreting gas from a warped disk which is fed by its
main sequence companion HZ Her. Similar warp disks have also been
proposed for SS 433 \citep{katz80,vd80} and V503 Cygni \citep{wb07}.
The origin of such warped disks is unclear, although there have been
many theories \citep{bp75,pet77,pp83,pl95,
  pt95,pringle96,od01}
.  Based on the models presented here, we suggest that the warp may
also be caused by the misalignment between the donor star's spin and
the binary system's orbit.  Although the 3D extension of these 2D
simulations is yet to be carried out, we anticipate the tilt of the
stellar spin axis may also be episodically modulated by internal
gravity waves.  One potential observational test is to search for
spin-orbit misalignment with the Rossiter-McLaughlin effect among
binary stars with relatively small $r$ (ie less affected by tidal
interactions) and relatively massive stars (which contain convective
cores and radiative envelopes).

3) {\it Classical Be stars} are rapidly rotating B stars which show
intermittent circumstellar decretion disks.  Approximately 12-17\% of
all B stars are Be stars.  There are many open questions with regard
to this class of stars (see \cite{portriv03} for a review).  First, it
is unclear how angular momentum is supplied to the gaseous disk in
order to expel it from the star.  Second, it is unclear why a large
fraction (although not all) of these stars rotate so rapidly (at
70-80\% critical).  IGW dynamics could contribute to both of these
outstanding questions.  First, the angular momentum required to launch
the equatorial shells of the star may come from IGW as originally
  suggested by \cite{ando86}.  The equatorial
region is precisely the region where we expect angular momentum
transport by IGW to be most effective and it may be that in a certain
parameter space IGW are even more efficient than we have discussed
here. Angular momentum transport by waves from the core to the surface
could cause the surface layers to spin extremely rapidly, causing
rapid rotation.  Our models do not show such extreme rotation rates,
but that could be a limitation of the numerics.  There is already some
observational evidence that IGW contribute to the Be mechanism.  Using
CoRoT observations \cite{nein12} have observed IGW which are not due
to $\kappa$ mechanism and therefore, most certainly IGW-inertial waves
driven by convection in the core and \cite{huat09} have detected
modulation of g-modes which are directly related to a Be outburst.

4) {\it Nitrogen enrichment in massive stars}.  Recent observations
\citep{Hunter2008} of massive stars found slowly rotating stars with
no surface nitrogen enhancement and rapidly rotating stars with strong
nitrogen enhancement, which are both well explained by models of
rotationally induced mixing.  However, a significant fraction of stars
cannot be explained by rotationally induced mixing alone.
\citet{Hunter2008} show that more than 10\% of their sampled stars are
enhanced in nitrogen without rapid rotation, while 20\% of their stars
are rapidly rotating without showing any significant nitrogen
enhancement. These stars contradict models which predict that
rotational velocity should be correlated with surface nitrogen
enhancement \citep{Brott2011}.  Therefore, these observations show
that other processes must be responsible for either the nitrogen
enhancement or the current rotation velocity. We propose that the
Nitrogen enhancement may indeed be due to rotational mixing but that
IGW can change the surface rotational velocity such that the observed
surface rotational velocity may not be a good indicator of the
rotational velocity of the whole star. For example, the surface may be
spun down due to IGW. In such stars, the star appears to be a slow
rotator but nitrogen has already been enhanced due to rotational
mixing in the past.  Similarly, IGW momentum transport may cause the
appearance of a rapid rotator in a system which has inefficient
rotational mixing and therefore, has yet to cause a nitrogen
enhancement.

\bibliographystyle{apj} \bibliography{hotjupob}
\section*{Acknowledgments}

We are grateful to G. Glatzmaier, K.B. MacGregor, Casey Meakin and
Eliot Quataert for helpful discussions.  We are thank an anonymous
referee for a detailed reading and suggestions which have improved
this manuscript.  Support for this research was
provided by a NASA grant NNG06GD44G, an NSF grant AST09-08807, a
UC/LAB fee grant and the Templeton Foundation.  T. Rogers is supported by an NSF ATM Faculty
Position in Solar physics under award number 0457631.  Computing
resources were provided by NAS at NASA Ames.

\clearpage

\clearpage

\clearpage
\clearpage
\clearpage
\clearpage
\clearpage

\clearpage
\clearpage
\clearpage
\clearpage
\clearpage
\clearpage
\clearpage

\clearpage
\end{document}